\documentclass[aps,pre, twocolumn, superscriptaddress, balancelastpage, showpacs, reprint,nofootinbib]{revtex4-2}

\usepackage[T1]{fontenc}
\usepackage{blindtext}
\usepackage{centernot}
\usepackage{graphicx}
\usepackage{amsmath,bbold}
\usepackage{times}
\usepackage{amssymb}
\usepackage{MnSymbol}
\usepackage{mathrsfs}
\usepackage{chemarr}
\usepackage{color}
\usepackage{url}
\usepackage{version}
\usepackage{mwe,tikz}
\usepackage[percent]{overpic}
\usepackage{bm}
\usepackage[export]{adjustbox}
\definecolor{linkcolor}{rgb}{0,0,0.6} 
\definecolor{forestgreen}{rgb}{0.13, 0.55, 0.13}
\definecolor{frenchblue}{rgb}{0.0, 0.45, 0.73}
\definecolor{burntsienna}{rgb}{0.91, 0.45, 0.32}
\usepackage{algorithm}
\usepackage[noend]{algpseudocode}
\usepackage{array}
\usepackage{multirow}

\usepackage{enumitem}
\usepackage{cases}
\usepackage[normalem]{ulem}
\usepackage{transparent}

\newcommand\rsout{\bgroup\markoverwith{\textcolor{red}{\rule[0.5ex]{2pt}{0.8pt}}}\ULon}

\newcommand{\nstar}{{n^\star}}

\newcommand{\dd}{\mathrm{d}}

\newcommand{\lambdaMax}{\lambda_\mathrm{Re}^\mathrm{max}}

\newcommand{\Real}{\mathrm{Re}}

\newcommand{\hatKdelta}{\hat K_\delta}
\newcommand{\Mstar}{M^\star}

\DeclareMathOperator\diag{diag}

\usepackage{lipsum}
\usetikzlibrary{patterns}

\newcolumntype{P}[1]{>{\centering\arraybackslash}p{#1}}
\newcolumntype{M}[1]{>{\centering\arraybackslash}m{#1}}

\usepackage[hidelinks, hyperfootnotes=false, colorlinks=true, linkcolor=black, citecolor=blue]{hyperref}

\begin{document}
  

\title{Patterns robust to Disorder in spatially-interacting Generalized Lotka-Volterra Ecosystems}

\author{Alessandro Salvatore}
\affiliation{LadHyX, CNRS, École polytechnique, Institut Polytechnique de Paris, 91120 Palaiseau, France}
\affiliation{Chair of Econophysics and Complex Systems, \'Ecole polytechnique, 91128 Palaiseau Cedex, France}

\author{Fabi\'an Aguirre-L\'opez}
\email{fabian.aguirre-lopez@ladhyx.polytechnique.fr}
\affiliation{LadHyX, CNRS, École polytechnique, Institut Polytechnique de Paris, 91120 Palaiseau, France}
\affiliation{Chair of Econophysics and Complex Systems, \'Ecole polytechnique, 91128 Palaiseau Cedex, France}

\author{Ruben Zakine}
\email{ruben.zakine@ladhyx.polytechnique.fr}
\affiliation{LadHyX, CNRS, École polytechnique, Institut Polytechnique de Paris, 91120 Palaiseau, France}
\affiliation{Chair of Econophysics and Complex Systems, \'Ecole polytechnique, 91128 Palaiseau Cedex, France}

\date{\today}

\begin{abstract}
How do interactions between species influence their spatial distribution in an ecosystem? To answer this question, we introduce a spatially-extended ecosystem of Generalized Lotka-Volterra type, where species can diffuse and interactions are nonlocal. We compute the criterion for the loss of stability of the spatially homogeneous ecosystem, and we show that the stability of the uniform state crucially depends on the most abundant species, and on the interplay between space exploration during one species generation and the interaction range. Focusing on the spectrum of the interaction matrix weighted by the species abundances, we identify a Baik-Ben Arous-Péché transition that translates into a transition in the final patterns of the species repartition. Finally assuming that the disorder is small, we exhibit an explicit solution of the dynamical mean-field equation for the species density, obtained as the fixed point of a nonlocal Fisher-Kolmogorov-Petrovski-Piskounov equation. Our work paves the way of future combined approaches at the frontier of active matter and disordered systems, with the hope of better understanding complex ecosystems like bacterial communities. 
\end{abstract}

\maketitle

\section{Introduction}

The first works on ecological models can be attributed to Alfred Lotka~\cite{lotka1920undamped} and Vito Volterra~\cite{volterra1926fluctuations} for independently proposing the mean-field equations that govern a two-species predator-prey system. As they stand, these equations are too simple to accurately describe ecosystems~\cite{maslov2017population,joseph2020compositional,dedrick2023does} but they highlight the possible endogenous population regulation dynamics. The Lotka-Volterra formulation can then be seen as the starting point to provide answers to important ecological questions, namely: Can one predict the extinction of a species? Can one foresee a species invasion? How does a species spread in a given environment?  
Answering these questions is extremely difficult in practice, since large scales experiments cannot be carried out easily. From a theoretical point of view, two approaches in physics have been followed in parallel to still bring insights to these questions.

A first approach builds on the seminal work of Fisher~\cite{fisher_wave_1937}, and Kolmogorov, Petrovski, Piskounov~\cite{kolmogorov1937}, and focuses on the spatial propagation of a small number of species, typically one, two, or three, subjected to diffusion, logistic growth and interaction~\cite{reichenbach2008self, tauber2024}. These models have proven efficient to understand bacteria spreading in controlled setups~\cite{deforet2019, lee2022}. Nonlocal interactions between species that can typically emerge from the sensing of a chemical in the medium can also be considered~\cite{pigolotti_species_2007, andreguetto_maciel_enhanced_2021,loreau2024Linking}.  

A second and parallel approach builds on the physics of disordered systems. More precisely, to avoid burdening biologists and ecologists, Robert May in 1972 addressed the question of the general stability of ecosystems in which the interactions between species are not known but are drawn from a probability distribution (say Gaussian), reflecting both the ignorance of interspecies interactions and the possible complexity and heterogeneity of interactions~\cite{may_will_1972}. Further using the tools of complex systems, this approach has known a recent upsurge of interest and has brought answers on the multiple dynamically accessible equilibria of an ecosystem~\cite{bunin_ecological_2017,galla_dynamically_2018, altieri_glassy2021, aguirre2024heterogeneous,poley2024interaction,park2024incorporating}, along with the dynamical transitions from stability to chaos and aging~\cite{baron_breakdown_2023, arnoulx_de_pirey_aging_prl2023,suweis2024generalized}, and the possible resilience of ecosystems when migration is possible~\cite{arnoulx_de_pirey_prx_2024,lorenzana2024, denk_tipping_2024}. 

In the article, we try to bridge the gap between these two approaches by considering a spatially-extended Lotka-Volterra ecosystem in which the interactions are random and nonlocal, a path that has been followed recently~\cite{olmeda2023long,loreau2024Linking, maritan_pnas2024}. After specifying the instability criterion of such ecosystems, we identify two regimes of pattern formation, dictated by a Baik-Ben Arous-Péché transition. In some regions of parameter space, we also find that the dynamics is controlled by a nonlocal Fisher-Kolmogorov-Petrovskii-Piskunov (F-KPP) equation, which is obtained by means of dynamical mean-field theory (DMFT). Using recent results on the F-KPP equation, we can obtain the stationary state of the spatially heterogeneous system. Our work constitutes a noticeable example where the non-trivial solution of a DMFT equation for a spatially-extended field can be written explicitly. Our findings are validated by extensive numerical simulations.

\section{A model with nonlocal interactions}

We consider $N$ species interacting in a $d$-dimensional domain $\Omega$, with $d=1$ or $2$. These species can diffuse in space. The abundance of each species is described by a field $\rho_i$ with $i\in\{1,\dots,N\}$, whose dynamics in space and time are given by the generalized Lotka-Volterra equations
\begin{align}
\begin{split}
    \partial_t\rho_i(x,t) =& D\nabla^2 \rho_i(x,t)\\
    &+\rho_i(x,t)\left( 1 +\sum_{j=1}^N W_{ij}K_\delta * \rho_j (x,t)\right),
    \label{eq:model_SEGLV}
    \end{split}
\end{align}
where we have introduced a diffusion coefficient $D$, an interaction kernel $K_\delta$ and coefficients $W_{ij}=-\delta_{ij}+A_{ij}$ with random $A_{ij}$ that translate the heterogeneity of possible interactions between different species. In this model, the reproduction rate and the carrying capacity are thus set to 1 for each species in absence of interactions. The heterogeneity in the diffusion can be considered as well and will be discussed later. The kernel $K_\delta\in L_2(\mathbb R^d)$ displays a typical interaction range of $\delta$ and we assume it can be cast into $K_\delta(x)=\frac{1}{\delta^d}Q(\frac{x}{\delta})$, with $\int_\Omega  Q (x) \dd x =1$,  $\int_\Omega   x Q (x)\dd x =0$, and $\int_\Omega   x^2 Q (x)\dd x <\infty$. The operator $*$ indicates the convolution in space domain, i.e. 
\begin{align}
    K_\delta * \rho(x,t)=\int_\Omega K_\delta(x-y)\rho(y,t) \dd y.
\end{align}
The interaction kernel translates the nonlocal interactions that emerge between species in a spatially-extended ecosystem. For instance, bacteria and fungi can interact through chemicals released in the medium and alter the replication process of other species.
Following the notations of \cite{galla_dynamically_2018,baron_breakdown_2023}, the coefficients $A_{ij}$ are defined by $A_{ii}=0$, and for any $i\neq j$, $A_{ij}$ are random and defined as
\begin{align}
    A_{ij} = \frac{\mu}{N} + \frac{\sigma}{\sqrt{N}} z_{ij},
\end{align}
where $\mu>0$ translates  the mean interaction type, cooperative ($\mu>0$) or competitive ($\mu<0$), $\sigma$ is the level of dispersion in the inter-species interactions, and the variables $z_{ij}$ are drawn from a Gaussian distribution with $\overline{z_{ij}}=0$, $\overline{z_{ij}^2}=1$ and $\overline{z_{ij}z_{ji}}=\gamma$. The overbar stands for the average over the Gaussian ensemble, and $\gamma\in[-1,1]$ indicates the correlation between coefficients, controlling the fraction of predator-prey interactions.

The main goal of the present article is to establish the phase diagram of the present model, notably specifying under which circumstances heterogeneous ecosystems remain stable, or in other words, that the population neither vanishes, nor diverges.

\section{How the most abundant species destabilizes the ecosystem}

\subsection{Homogeneous densities as a starting point}

We want to know under which conditions the spatially homogeneous population is no longer stable. This criterion can be found by a direct linear stability analysis on Eq.~\eqref{eq:model_SEGLV}. First, one should determine the homogeneous fixed point of the dynamics, i.e. the state $\bm{n}= (n_1,\dots,n_N)^\top$ of spatially homogeneous densities $n_i$ that satisfy the stationary equations 
\begin{align}
    0 =n_i (1-n_i+ \sum_{j=1}^N A_{ij}n_j).
\end{align}
The notation $n_i$ will refer to the density or abundance of the species $i$ in the zero-dimensional system. The distribution of these fixed points \textit{reached dynamically} has already been the object of intense research~\cite{galla_random_2006, bunin_ecological_2017, altieri_properties_2021,arnoulx_de_pirey_aging_prl2023,garnier2021new}, and is already a nontrivial question. In essence, since one looks for positive (or null) densities, the fixed point equation yields $n_i=0$ or $\sum_{j=1}^N (1-A_{ij})n_j=1$, which, assuming $W\equiv -I+A$ is invertible, leads to 
\begin{align}
    n_i = -\sum_{j=1}^N W^{-1}_{ij}=\sum_{j=1}^N (I-A)^{-1}_{ij}.
\end{align}
If the right-hand side is positive for all $i$, then the ecological system is said \textit{feasible}~\cite{bunin_ecological_2017, galla_dynamically_2018}.
If the right-hand side is negative for some $i$, it means that the targeted fixed point cannot be reached by the dynamics, and one should discard the species that is extinct ($\rho_i\leq 0$) from the equation. In that case, the system is said \textit{non-feasible}. One must pay attention to the fact that discarding an extinct species $i$ in the dynamics is equivalent to setting $W_{ij}=W_{ji}=0$ for all $j$. Suppressing the line and column $i$ in $W$ yields a new interaction matrix $Y$ whose coefficients can be strongly correlated~\cite{baron_breakdown_2023}. In practice, the reduced matrix can also be obtained numerically with very high accuracy by running a dynamics without space, i.e.
$\dd n_i(t)/\dd t=n_i(t)(1+\sum_{j=1}^NW_{ij}n_j(t))$,
let the dynamical system evolve and reach a fixed point, and discard the coefficients in $W$ corresponding to species whose densities are $0$ up to numerical error, typically $\sim 10^{-12}$.

In what follows, we will denote $R$ the diagonal matrix such that $R_{ii}=n_i$ for all $i$, and we will derive the instability criterion for the extended ecosystem. 

\subsection{Criterion for instability}
\label{sec:criterion_for_instability}

We assume that the ecosystem is a priori feasible, i.e $R_{ii}>0$ for all $i$. The derivation for a non-feasible ecosystem is more involved and is provided in Appendix~\ref{app:non_feasible}. It will however lead to the same instability criterion. We focus on the evolution of a perturbation of the density. The perturbation is denoted $\bm \psi(x,t)=(\psi_1,\dots,\psi_N)^\top$ and we thus write $\bm \rho(x,t)=\bm{n}+\bm \psi(x,t)$. Linearizing Eq.~\eqref{eq:model_SEGLV} close to the feasible equilibrium yields
\begin{align}
    \partial_t \bm \psi(x,t) = D\partial_x^2 \bm \psi(x,t) + RWK_\delta* \bm \psi(x,t).
\end{align}
In Fourier space, where we denote by $\hat \psi(k)=\int_\Omega \psi(x) e^{i kx}\dd x$ the Fourier transform of a function $\psi$, we obtain the evolution of a mode $k$:
\begin{align}
    \partial_t \bm{ \hat\psi}(k,t) = \left(-Dk^2 I+ \hat K_\delta(k) RW \right) \bm{\hat\psi}(k,t).
\end{align}
We denote by $\lambda(M)=\{\lambda_i(M)\}_{i=1,\dots,N}$ the spectrum of a matrix $M$. The homogeneous fixed point loses its stability when one of the eigenvalues of the matrix $\Lambda(k)=-Dk^2I+\hat K_\delta(k)RW$ has a positive real part. Since $-Dk^2 I$ is diagonal, one is thus left with the computation of the spectrum $\lambda(RW)$. It turns out that the spectrum of the matrix $RW$ has been studied in~\cite{ahmadian2015,stone_feasibility_2018,baron_breakdown_2023}, and always lies in the $\Real<0$ part of the complex plane for feasible ecosystems. From this, we conclude that if $\hat K_\delta(k)\geq 0$, then $\lambda_i[\Lambda(k)]\leq 0$ and the feasible ecosystem remains stable. We will thus assume in what follows that the Fourier transform of the kernel $K_\delta$ can be negative. Since we are looking for $\lambda_\mathrm{max}[\Lambda(k)]$, the instability will come from the eigenvalue of $RW$ with the smaller (or most negative) real part, that we will denote $\lambda_\mathrm{min}(RW)$.
This onset of instability is here defined by the manifold where for the marginally stable mode $k_c$, one has $0=\lambda_\mathrm{max}[\Lambda(k_c)]=\partial_k \lambda_\mathrm{max}[\Lambda(k_c)]$. Using the fact that $\hat K_\delta(k)=\hat Q(\delta k)$, the marginally stable mode $k_c$ can be written as $k_c=\zeta_c/\delta$, with $\zeta_c$ the smallest positive solution of
\begin{align}
    \frac{\hat Q'(\zeta)}{\hat Q(\zeta)} \zeta=2.
    \label{eq:definition_k_c}
\end{align}
The pattern wavelength is thus independent of the surviving species characteristics, or of the diffusion coefficient, but simply dependent on the kernel $Q$ and on the interaction length $\delta$.
Finally injecting $k_c=\zeta_c/\delta$ in the instability condition $0=-Dk_c^2+\hat K_\delta(k_c)\lambda_\mathrm{min}(RW)$, we find that the homogeneous system becomes unstable when the ratio $D/\delta^2\equiv \alpha$ is smaller than a critical value denoted $\alpha_c$ and defined by
\begin{align}
    \alpha_c = \frac{\hat Q(\zeta_c)}{\zeta_c^2}\lambda_\mathrm{min}(RW).
        \label{eq:critical_D_vs_max_lambda}
\end{align}
Recalling that the reproduction rate was set to $1$, we understand that $\sqrt{D}/\delta$ is the ratio between two quantities: the typical diffusion length during one species generation, and the interaction range $\delta$. If the ratio is large enough, interactions become purely local, and since all species have the same diffusivity, the behavior is dictated by its 0-dimensional counterpart. If the ratio is small, distant interactions influence reproduction faster than species displacement, possibly leading to constructive or destructive reactions, hence an instability at finite wavelength, depending on the kernel shape.

\begin{figure}
    \centering
    \includegraphics[width=1\linewidth]{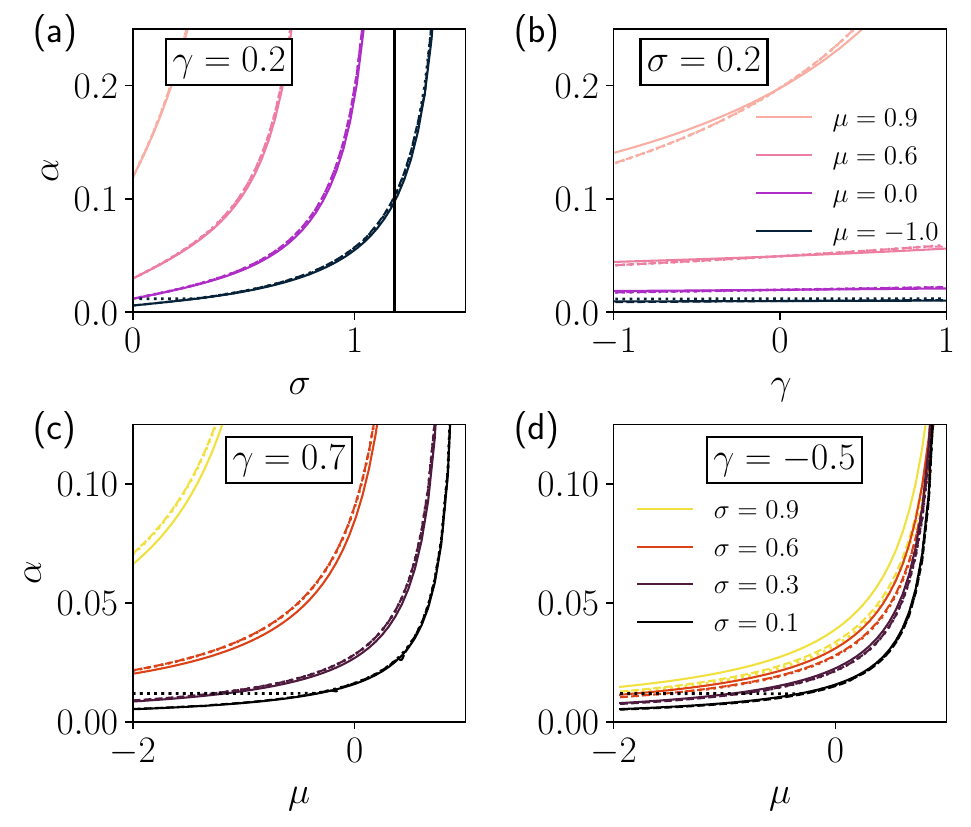}
    \caption{Transition lines $\alpha_c(N,\sigma,\mu,\gamma)$. Solid lines are plotted from perturbation theory, using Eqs.~\eqref{eq:critical_D_vs_max_lambda} and \eqref{eq:lambda_min_naive}.  Dashed lines are obtained via a DMFT computation, see Appendix~\ref{app:dmft_theory}. Dotted lines are obtained via random matrix theory (RMT), see Eqs.~\eqref{eq:lambda_min_rmt_maintext}, \eqref{eq:lambda_min_bulk}, \eqref{eq:lambda_min_outlier}. The RMT and the DMFT predictions overlap almost perfectly, except in the regions where the spectrum of $RW$ displays an outlier eigenvalue that determines the stability of the system. For all predictions, Eq.~\eqref{eq:max_rho_scaling} has been used to assess $\max_i n_i$.
    Parameters: $N=200$ in (a,b,c), other specific values are given in the panels. Panels (a) and (b) share the same values for $\mu$, while panels (c) and (d) share the values of $\sigma$. The interaction kernel here is $K_\delta(x)= 1/(2\delta)$ for $|x|<\delta$, and $0$ otherwise. 
    The system is homogeneous in the region $D/\delta^2\equiv\alpha>\alpha_c$. For $\alpha<\alpha_c$, density profiles display spatial modulation with main frequency $k_c$. The vertical black line in panel (a) is at $\sigma_c=\sqrt{2}/(1+\gamma)$ and indicates the loss of stability from the mode $k=0$. For $\sigma>\sigma_c$ the zero-dimensional dynamics is chaotic.}
    \label{fig:transition_lines}
\end{figure}

To compute the spectrum of the matrix $RW$, we use perturbation theory, \textit{a priori} restrained to $|\mu|$, $\sigma\ll 1$. We will see in Fig.~\ref{fig:transition_lines} that it leads to a very good approximation of the instability onset.
We start by remembering that $W=-I+A$, with $\|A\|_\infty\equiv \max_{ij}A_{ij}=O(\sigma)+O(\mu)$ a perturbation. In the canonical basis, using the fact that $A_{ii}=0$ for all $i$, perturbation theory yields
\begin{align}
    \lambda_i(RW)=-n_i+O(\sigma^2)+O(\mu). 
    \label{eq:lambda_order1}
\end{align}
One cannot compute the higher order terms of the expansion via perturbation theory because the distance between the unperturbed eigenvalues is typically much smaller than the amplitude $\sigma$ of the perturbations ($\min_{ij}|n_i-n_j|=O(\sigma/N)\ll \sigma$).
The minimum eigenvalue of $RW$ is thus given by
\begin{align}
    \lambda_\mathrm{min}(RW) \simeq - \max_{1\leq i\leq N} n_i.
    \label{eq:lambda_min_naive}
\end{align}
In this system, interestingly enough, the stability is thus fully determined by the behavior of the most abundant species, a feature that could be measured and tested in controlled ecosystems. Also, this result is valid for any correlation $\gamma$, assuming the zero-dimensional dynamics has converged to a feasible state. Finally, the condition $\hat K_\delta(k)\leq 0$ seems peculiar because it means that the specific shape of the kernel plays a role, and as such, endangers the very notion of universality. The sufficient and necessary conditions to obtain negative Fourier transforms of a distribution remain poorly understood physically~\cite{tuck_positivity_2006, giraud_positivity_2014}. The role of the kernel had already been unveiled in a similar model~\cite{pigolotti_species_2007}, without diffusing species though. Its effect on the possible resulting patterns has also been explored in~\cite{andreguetto_maciel_enhanced_2021}. Note finally that the mechanism that leads to an instability here is also different from a Turing pattern formation, which emerges in a reaction-diffusion system when species do not have identical diffusivity~\cite{turing1990chemical}. 

The minimum eigenvalue of $RW$ can actually be obtained more accurately by means of random matrix theory. In Appendix~\ref{app:rmt_results}, we show that the correction to the 0th order reads
\begin{align}
\lambda_\mathrm{min}(RW) = \min\left\{ \lambda_\mathrm{min}^\mathrm{bulk}, \lambda_\mathrm{out}  \right\},
\label{eq:lambda_min_rmt_maintext}
\end{align}
with 
\begin{align}
\lambda_\mathrm{min}^\mathrm{bulk} = -n_{\mathrm{max}} \left(\frac{1}{2}+\frac{1}{2}\sqrt{1+\frac{4M\gamma\sigma^2}{n_{\mathrm{max}}-M}} \right),
\label{eq:lambda_min_bulk}
\end{align}
and 
\begin{align}
 \lambda_\mathrm{out} = -1 +\frac{\gamma \sigma^2}{\mu(1-\mu)},
 \label{eq:lambda_min_outlier}
\end{align}
and where $M=\frac{1}{N}\sum_{i=1}^N n_i$ denotes the mean abundances of species, and $n_\mathrm{max}=\max_{1\leq i\leq N} n_i$. These expressions where obtained using a mean-field approximation, and assuming that $\sigma^2<|\mu|$ to compute the outlier.

We then use the fact that $\max_i n_i$ can be obtained in the large $N$ limit. Indeed, the $n_i$ are shown to be drawn from a truncated Gaussian distribution in the thermodynamic limit. This distribution can be obtained by means of dynamical mean-field theory (DMFT) computations and we refer to Appendix~\ref{app:zero_dim_results} for a summary of the results derived in Ref.~\cite{bunin_ecological_2017,galla_dynamically_2018}. Using the fact that the maximum value of $N$ draws of a Gaussian random variable with mean $0$ and variance $v$ scales as $\sqrt{2v\log N}$, we obtain for $n_i$:
\begin{align}
    \max_i n_i = \frac{1+\mu M+\sigma \sqrt{2q\log N}}{1-\gamma \sigma^2 \chi},
    \label{eq:max_rho_scaling}
\end{align}
where we have injected the DMFT prediction of the first and the second moment of the fixed-point distribution, denoted $M$ and $q$, respectively, and the response coefficient $\chi$ to obtain this result. 

All in all, the predictions of RMT and the ones from perturbation theory are remarkably close, as observed  in Fig.~\ref{fig:transition_lines}. They agree even for large values of $\mu$ and $\sigma$, if there is no outlier eigenvalue. The results from RMT are also confirmed by a linear stability analysis on the dynamical mean-field equations, see Appendix~\ref{app:dmft_theory}.

Finally, it appears that $\max_i n_i$ diverges as $\sqrt{\log N}$ when $N\to\infty$. When looking at  the expression of $\alpha_c$, we conclude that whenever $\hat K_\delta(k)<0$, a homogeneous system will always loose stability via the growth of the critical mode $k_c$ as the number of species in the ecosystem increases. Hence, in the following we will no longer assume that $N\to\infty$ that would trivially lead to instability but rather consider $N$ large but finite.

\subsection{Phase diagram}
\label{sec:patterns_and_phase_diag}

Having identified the instability criterion, we now turn to the phase diagram and we describe the new phases that emerge when the system is no longer homogeneous.

We work at a finite $N$. In that case, the instability can either come from the mode $k_c$ or from the mode $0$ depending on the values of $\alpha$, $\sigma$, $\gamma$ and $\mu$. The various transition lines are displayed in Fig.~\ref{fig:transition_lines}, for a given interaction kernel $K_\delta$. If the instability comes from the mode $0$, the dynamics of the system will follow the zero-dimensional one: Either the average abundance $M$ may diverge without the loss of stability of the homogeneous fixed point, or the fixed point loses its stability, multiple attractors appear and a chaotic dynamics is expected~\cite{bunin_ecological_2017, galla_dynamically_2018}. Interestingly enough, the simulations show that the densities become spatially uniform in the chaotic phase. This behavior is not surprising since the time scales of the population dynamics diverge with aging in the chaotic phase~\cite{arnoulx_de_pirey_aging_prl2023, arnoulx_de_pirey_prx_2024}, but the time scale of diffusion remains finite, hence the spatial homogenization of the densities. 
We will assume that the instability comes from the mode $k_c$.
\begin{figure}
    \centering
    \includegraphics[width=1\linewidth]{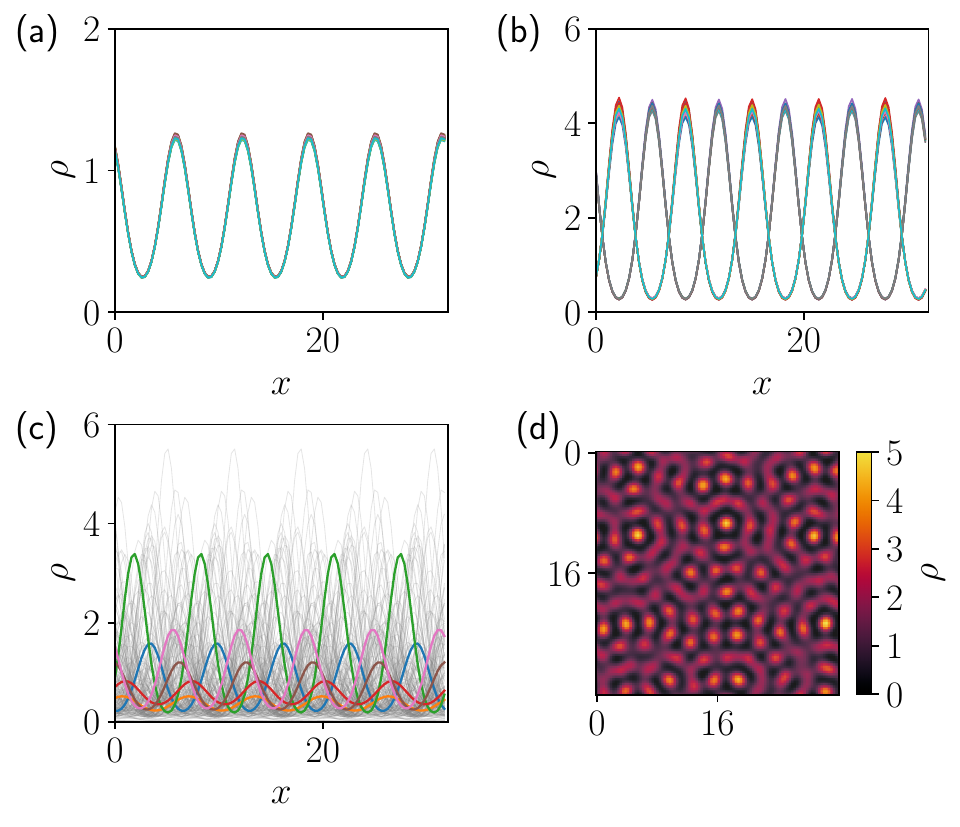}
    \caption{(a) and (b) show a few density profiles in $d=1$, for small disorder $\sigma=0.01$, and $\mu=-0.5$ in (a), while $\mu=0.4$ in (b). Panel (c) displays a state of the system when the disorder $\sigma=0.5$, with a few species in color lines out of the $N=200$ in the system, represented in grey.
    (d) An average density profile $\rho(x)=\frac{1}{N}\sum_{i=1}^N \rho_i(x) $ in $d=2$ in panel. 
    Parameters: $L=32$, $L_\mathrm{grid}=100$, $\delta=4$. For (a), $N=200$, $\gamma=0$, $D=0.15$. For (b) $N=200$, $\gamma=0$, $D=0.2$. For (c) $\mu=-0.5$, $D=0.15$. For (d) $N=50$, $\sigma=0.01$, $\mu=0.4$, $\gamma=0$.}
    \label{fig:4patterns}
\end{figure}

We then perform extensive numerical simulations to explore the phase diagram. In practice, and to keep a reasonable convergence time, we solve the dynamics via a semi-spectral scheme on a one-dimensional domain with periodic boundary conditions, for $N\geq 200$.
We first confirm that for $\alpha>\alpha_c$, the homogeneous state remains stable, see symbols $(\circ)$ Fig.~\ref{fig:phase_diag_phi_M_vs_sigma}(a) and (d). We then identify 3 regimes when the homogeneous state loses stability via $k_c$, i.e. when $\alpha<\alpha_c$.
(i) For small values of the dispersion $\sigma$, the system can reach a stationary state where all the species densities display spatial modulation of wavelength $2\pi/k_c$, as displayed in Fig.~\ref{fig:4patterns}(a) and (b).
(ii) For larger values of $\sigma$ and $\alpha$, the density profiles keep the patterned structure but are evolving with time. They typically oscillate, and the species propagate in the medium, without displaying mass explosion, see Fig.~\ref{fig:4patterns}(c).
(iii) For even larger values of $\sigma$, the instability at $k_c$ leads to exponential divergence of the pattern amplitudes and the abundances explode, see symbols ($\times$) in Fig.~\ref{fig:phase_diag_phi_M_vs_sigma}(a) and (d).

Our numerical findings are consistent with our theory. They are also in agreement with the linking between scales observed numerically in~\cite{loreau2024Linking}, since our variable $\alpha$ shows that the system will pattern when diffusion $D$ is small enough or when the range of interaction $\delta$ is large enough.

\subsection{Surviving fraction and abundances}

\begin{figure}
    \centering
    \includegraphics[width=1\linewidth]{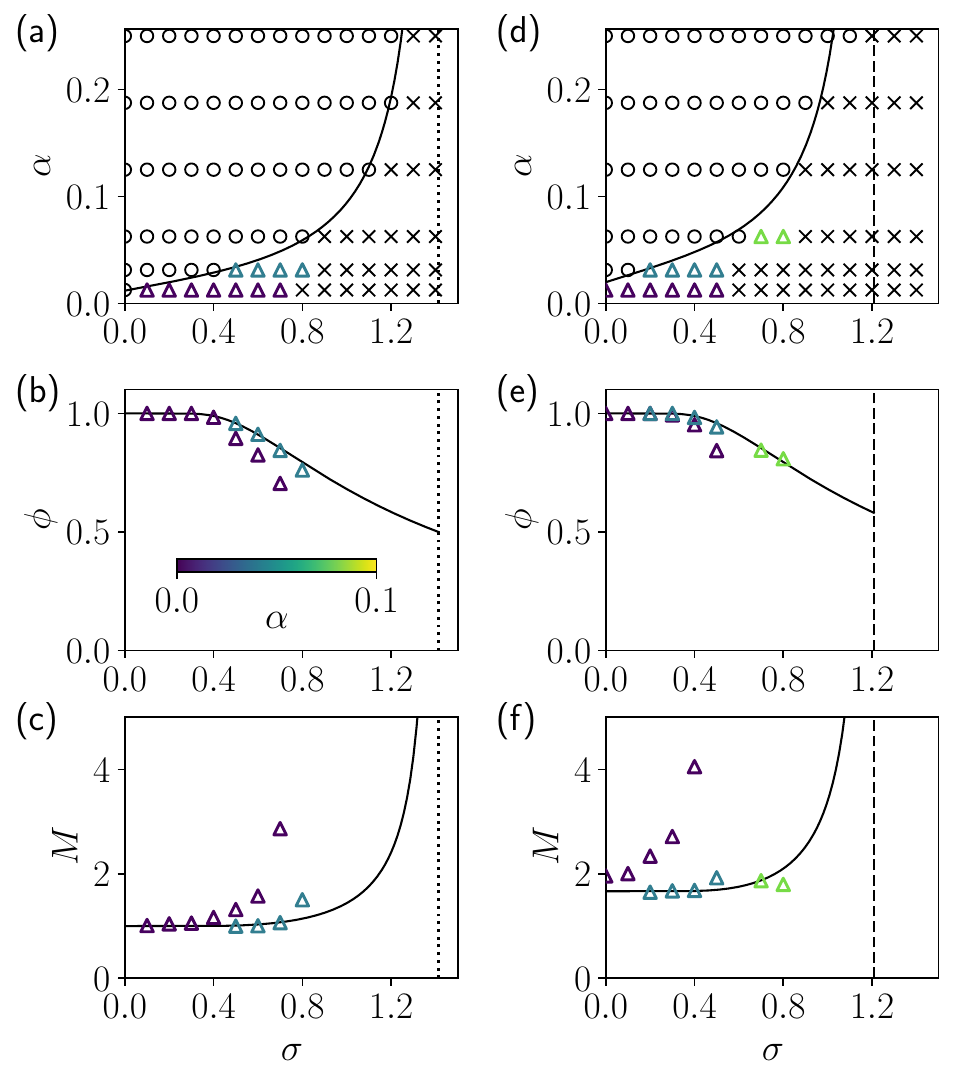}
    \caption{Phase diagram, surviving fraction $\phi$, and mean abundances $M$ as a function of the dispersion $\sigma$, for $\mu=0$ in (a,b,c), and $\mu=0.4$ in (d,e,f). Symbols: ($\circ$) flat density profiles, ($\bigtriangleup$)~patterns, $(\times$)~diverging abundances. The solid line in (a) and (d) indicates the destabilization of the homogeneous phase by the mode $k_c$.
     The solid lines in (b,c,e,f) indicate the values in the zero-dimensional system. The vertical dotted line in (a,b,c) indicates the destabilization of the zero-dimensional system to the chaotic phase. The vertical dashed line in (d,e,f) indicates the frontier of diverging abundances in the zero-dimensional system. When the density profiles are homogeneous, the surviving fraction $ \phi$ and mean abundance $M$ lie exactly on the zero-dimensional prediction curve, and are thus not displayed.
     Other parameters: $\gamma=0$, $\delta=4$, $N=200$, system size $L=32$. Each point in these panels is obtained by averaging over $4$ independent simulations.}
    \label{fig:phase_diag_phi_M_vs_sigma}
\end{figure}
An important question that initially motivated this work is to know whether the additional spatial dimension may prevent species from becoming extinct. We hypothesized that moving in space could allow a species to escape a predator or to escape the competitive interactions on a given site. To measure the effect of space on the ecosystem, we rely on two observables, namely, the surviving fraction $\phi$ and the mean abundance $M$. In the system, a species is considered surviving if its mean abundance over space denoted $\langle\rho_i\rangle_x$ is larger to some threshold, typically $10^{-8}$. The surviving fraction $\phi$ is the ratio of the number of surviving species denoted $N_S$ and the total number of species $N$.

Interestingly enough, we find that the surviving fraction when species can spread spatially is not higher than the surviving fraction of the zero-dimensional system, when species are forced to interact in the same point of space. This is shown in Fig.~\ref{fig:phase_diag_phi_M_vs_sigma}(b) and (c) where the results of the simulations closely follow the prediction of $\phi$ in a zero-dimensional system. Even in the case of cooperative interactions ($\mu>0$), the surviving fraction is bounded by the 0d prediction. For very low $\alpha$ the surviving fraction is even smaller than the 0d prediction.
On the other hand, the mean abundance $M$ of the species across space, defined as $M=1/N\sum_{i=1}^N \langle \rho_i\rangle_x$ is found to be significantly larger than the 0d prediction for small $\alpha$, see Fig.~\ref{fig:phase_diag_phi_M_vs_sigma}(c) and (f).
As $\sigma$ increases, the abundances eventually diverge while the density profiles keep their spatially-periodic structure. 
All in all, our results indicate that space allows for a diverging biomass for lower levels of heterogeneity in the interactions. This fact contrasts with those of \cite{lorenzana2024,denk_tipping_2024}. Even though survival fraction increases with diffusion (higher values of $\alpha$), in our case average mass, $M$, decreases for larger diffusion, as it can be seen in Fig.~\ref{fig:phase_diag_phi_M_vs_sigma}.

\section{F-KPP equations and stable patterns}

In the previous section, we have shown  that the flat solution was destabilized for $\alpha<\alpha_c$. However, a proof is missing confirming the convergence to some state, stationary or not, in which the species densities remain bounded. We address this question below.

\subsection{The limit case $\sigma=0$}
Confirming the existence of a patterned solution to a partial differential equation can usually be done via the amplitude equation describing the evolution of the large wavelength modulations of the sinusoidal patterns emerging at the critical wavelength $k_c$. For $\sigma>0$, the disorder on the interacting coefficients can lead to oscillations and non-stationary solutions, as shown by the PDE solutions.  To get insights on the dynamics and to capture the transitions that we observe, we derive the DMFT equation describing the typical behavior of a random species in the ecosystem. This derivation is carried out in Appendix~\ref{app:dmft_theory}.
Finding a complete ansatz for the DMFT equation is usually out of reach in the general case. For $\sigma=0$ however, the disorder only lies in the initial conditions, and the DMFT equation can be strongly simplified, as the self-consistent noise and the response term vanish. The dynamics of a typical species thus reads
\begin{align}
\begin{split}
        \partial_t \rho(x,t)=& D\nabla^2 \rho(x,t)\\ &+ \rho(x,t)K_\delta*\left[1-\rho(x,t) + \mu M(x,t) \right].
        \label{eq:dmft_equation_sigma0}
\end{split}
\end{align}
Since the noise has vanished, all species now follow the same evolution equation. If $\mu=0$, species are not interacting with each other, and their final density profile will be identical, up to some phase shift that depends on the initial conditions. In particular, each species density satisfies a nonlocal Fisher-Kolmogorov-Petrovskii-Piskunov (F-KPP) equation~\cite{berestycki_non-local_2009,achleitner_bounded_2015, kuehn_validity_2017}, whose stationary fixed point is either homogeneous or patterned~\cite{faye_modulated_2015}. In $d=1$, the stationary patterns span on the whole spatial domain and are pure sinusoids close to the onset, and in $d=2$ a species condenses in a triangular lattice. For $\mu\neq0$, the different species do interact and we can assume that the initial conditions will not be relevant anymore to determine the density profile at long times. Numerically, we observe two distinct behaviors, depending on the sign of $\mu$. 

For $\mu<0$ (competitive interactions), the species end up all overlapping into a single stationary patterned state, similarly to what is shown in Fig.~\ref{fig:4patterns}(a) in $d=1$, or on the same triangular lattice in $d=2$ (not shown). In that case, the species profiles cannot be distinguished from the mean $M(x,t)$. If one assumes that, indeed, each profile can then be expanded as $\rho(x,t)=M(x,t)+o(|M(x,t)|)$, one obtains to leading order the dynamics for the mean: 
\begin{align}
\begin{split}
        \partial_t M(x,t)=& D\nabla^2 M(x,t) \\
        &+ M(x,t)K_\delta*\left[1 -(1-\mu) M(x,t) \right],
        \label{eq:dmft_equation_sigma0_mean}   \end{split}
\end{align} 
and we recover a nonlocal F-KPP equation, now satisfied for the mean $M(x,t)$. 
A solution to this equation is a stationary patterned profile, that we will denote $M^\star(x)$, for which densities remain bounded. In our case $\sigma=0$, close to the onset of patterning, the modes of interest are all around $k_c$~\cite{faye_modulated_2015}, where $k_c$ is defined by Eq.~\eqref{eq:definition_k_c}. 

\begin{figure}
    \centering
    \includegraphics[width=1\linewidth]{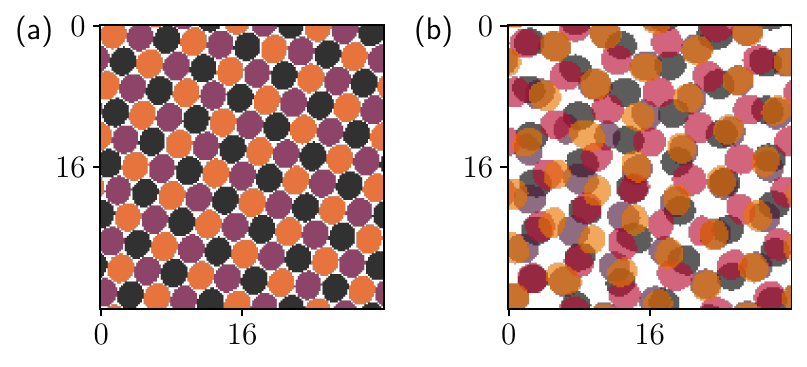}
    \caption{Two-dimensional arrangements of species when $\sigma=0$ and $\mu=0.4$, starting from homogeneous profiles. Panel (a): solution of the PDEs for $N=3$ interacting species. Panel (b): 4 random species out of the $N=50$ interacting in the system. Each color represents a different species, and only the regions where a species density is larger than some threshold $a$ (here $a=2$) is colored. In both systems, a species mostly occupies the nodes of a triangular lattice. The interaction kernel here is $K_\delta(x)= 1/(\pi\delta^2)$ for $|x|<\delta$, and $0$ otherwise.  Other parameters: $L=32$, $\delta=4$, $L_\mathrm{grid}=150$, $\Delta t=0.1$. }
    \label{fig:2d_patterns_only}
\end{figure}

For $\mu>0$ (cooperative interactions), the species spread spatially and self-organize into two groups (in $d=1$) with identical static pattern profiles (one group density is shifted by half a period with respect to the other), similarly to what is shown in Fig.~\ref{fig:4patterns}(b). In $d=2$ dimensions, group repulsion is also found but the species may end up in a frustrated state. Indeed, each species arranges in a triangular lattice but these lattices repel each other. For 3 species only, the lattices do not overlap because each species can occupy the nodes of a different triangular lattice, see Fig.~\ref{fig:2d_patterns_only}(a). Once a fourth species is added, the new triangular lattice of this species is repelled by the 3 others. As the number of species increases, the frustration enhances the apparition of topological defects that suppress the long-range translational and rotational orders, see Fig.~\ref{fig:2d_patterns_only}(b). In Fig.~\ref{fig:4patterns}(d), we display the average density of species in this frustrated system.

The drastic change of behavior as $\mu$ changes sign is the signature of a phase transition. We relate the transition in the patterns to a transition in the spectrum of the matrix $RW$. We sketch the argument below. For $\sigma=0$, when $\mu<0$ the minimum eigenvalue of the matrix $RW$ is $-1$ and is of multiplicity $1$, while for $\mu\geq0$, $\lambda_\mathrm{min}(RW)=1/(\mu-1)$ and is of multiplicity $(N-1)$, as detailed in Appendix~\ref{app:rmt_results}. The value $\mu=0$ corresponds to the outlier crossing the edge of the bulk of the spectrum (although degenerate for $\sigma=0$), a second-order transition referred to as the Baik–Ben Arous-Péché (BBP) transition~\cite{baik_ben-arous_2005}. Focusing on the case $\mu<0$, one can show that the eigenvector related to the outlier eigenvalue $-1$ is $R(1,\dots,1)^\top=\frac{1}{1-\mu}(1,\dots, 1)^\top$. The function $M(x,t)$ is then given by $\frac{1}{N}(1,...,1)^\top\bm \rho(x,t)$, the projection of the density vector $\bm \rho(x,t)$ on the only unstable direction $(1,\dots,1)^\top$ that destabilizes all the species in the same direction. The species finally align and the only dynamically relevant equation is indeed the equation of the mean $M(x,t)$. For $\mu>0$, there is no longer a unique unstable direction, and all the dynamically relevant directions (and equations) are to be considered, which translates in the PDE solution into several groups of interacting patterns. 

It may appear paradoxical that the species split in several families when interactions are cooperative, while they overlap when they compete. We rationalize this behavior \textit{a posteriori} when we compare the interaction radius $\delta$ of the kernel (here a step) to the period of the patterns $\lambda_p$. In particular, focusing in $d=1$ for simplicity, we have $\delta<\lambda_p<2\delta$, which means that a species, labeled $A$, that condenses at the maximum density of the patterns of a second one, labeled $B$, will not interact with the other density peaks of the species $B$, a favorable situations when species compete. Shifting the settlement of species $A$  by half of a period allows the species $A$ to interact with two density peaks of $B$, beneficial for abundances when interactions are cooperative. The phenomenon described here is possible because the kernel somehow weights distant interactions as relevant as local ones, which is tightly related to the fact that its Fourier transform has negative values. And indeed, we have checked that this behavior is recovered with other kernels whose Fourier transforms display negative values.
Finally, taking a nonzero $\sigma$ such that an outlier eigenvalue can still exist in the spectrum should not change the global picture of the pattern transition. This is the subject of the next section.

\subsection{Expansion close to $\sigma=0$}

\subsubsection{Case $\mu<0$: instability from the outlier eigenvalue}

What is the level of dispersion in the interactions for which the patterns are stable? 
We have shown in the previous paragraph that for $\sigma=0$, one can recover the nonlocal F-KPP equation satisfied by the mean $M(x,t)$. At the onset of patterning, we have shown that the solution $\Mstar(x)$ is spatially oscillating with wavenumber $k_c$, and that the patterns are static and stable. Our goal is now to expand for $\sigma$ small but nonzero.

\begin{figure}
    \centering
\includegraphics[width=1\linewidth]{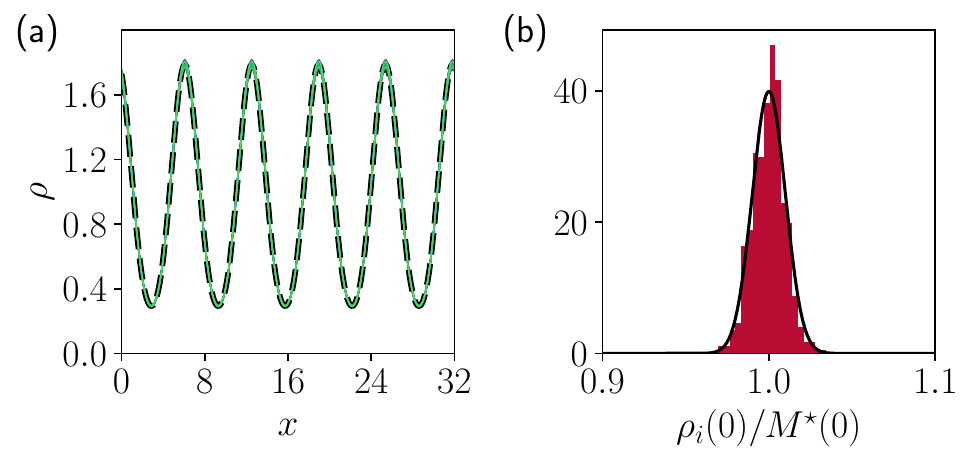}
    \caption{(a) A few density profiles (color lines), and mean profile $\Mstar(x)$ (black dashed line). (b) Rescaled distribution of abundances $\rho_i(0)/\Mstar(0)$ that fits the predicted Gaussian distribution of mean 1 and variance $\sigma$, see Eq.~\eqref{eq:dmft_expansion_distribution}. Parameters: $\sigma=0.01$, $\mu=-0.1$, $\gamma=-0.7$, $\delta=4$, $D=0.15$, $N=500$, system size $L=32$, number of gridpoints $L_\mathrm{grid}=200$. We checked that identical results are obtained for $\gamma=0$ and $\gamma=0.7$, which confirms that $\gamma$ does not play a role to leading order in $\sigma$.}
    \label{fig:profiles_and_dispersion_dmft_expansion}
\end{figure}

We start back from Eqs.~\eqref{eq:lambda_min_naive}
and \eqref{eq:max_rho_scaling}, and we rescale $\sigma\to \sigma'/\sqrt{\log N}$, such that the quantity $\sigma \max_{1\leq i\leq N} n_i$ stays of order $1$, and the critical line in parameter space $(\sigma',\alpha)$ is not collapsing on the $y$ axis in the limit $N\to\infty$.
In that limit, the variance of the distribution of fixed points goes to zero, and the mean is $M=1/(1-\mu)$.
We are going to expand close to the $\Mstar(x)$ solution. We have for some random species $\rho(x,t)$:
\begin{align}
    \rho(x,t) =M^\star(x) +\frac{\sigma'}{\sqrt{\log N}} \psi(x,t),
    \label{eq:expansion_rho_M_psi}
\end{align}
with $\psi(x,t)$ a species dependent field. We use the fact that $\Mstar(x)$ is solution of \eqref{eq:dmft_equation_sigma0_mean}, and retaining leading order terms in the DMFT equation yields
\begin{align}
\begin{split}
    \partial_t \psi(x,t) =& D\nabla^2\psi(x,t) \\
    &+ \psi(x,t)K_\delta*\left(1-M^\star(x)
    + \mu M^\star(x)\right)\\ 
    &+ M^\star(x)K_\delta*\left(-\psi(x,t) +  \eta(x,t)\right).
    \end{split}
    \label{eq:expanded_DMFT}
\end{align}
The DMFT noise $\eta(x,t)$, that satisfies $\langle \eta(x,t) \eta(x',t')\rangle = \langle \rho(x,t)\rho(x',t')\rangle$, needs simply to be evaluated to leading order. The correlation now reads
\begin{align}
    \langle \eta(x,t) \eta(x',t')\rangle
    &= \langle \Mstar(x)\Mstar(x')\rangle +O\left(\sigma\right)\\
    &=  \Mstar(x)\Mstar(x'),
\end{align}
to leading order. The correlation is no longer a function of $t$ and $t'$, so it is constant in time.
In space, the correlation is now factorized, which means that the covariance matrix is of rank 1. Hence, the noise $\eta(x,t\to\infty)$, that is Gaussian, is now deterministic, once $\eta(x=0,t\to\infty)$ given. We drop the time dependence and we have
\begin{align}
    \eta(x) = \frac{\eta(0)}{\Mstar(0)}\Mstar(x)\equiv \xi \Mstar(x),
\end{align}
with $\xi$ a Gaussian random variable of mean 0 and variance 1.
Injecting in Eq.~\eqref{eq:expanded_DMFT}, we have
\begin{align}
\begin{split}
    \partial_t \psi(x,t) =& D\nabla^2\psi(x,t) \\
    &+ \psi(x,t)K_\delta*\left(1-M^\star(x) + \mu M^\star(x)\right) \\
    &+ M^\star(x)K_\delta*\left(-\psi(x,t) +  \xi\Mstar(x)\right),
    \end{split}
\end{align}
which accepts $\psi(x,t) =\xi\Mstar(x)$ as a static solution. We are now able, using Eq.~\eqref{eq:expansion_rho_M_psi}, to give an explicit formula for the density profiles close to $\sigma=0$:
\begin{align}
    \rho(x) = \Mstar(x)\left(1 +\frac{\sigma'}{\sqrt{\log N}}\xi\right).
    \label{eq:dmft_expansion_distribution}
\end{align}
We have thus explicitly shown that the patterns survive a small amount of disorder in the interaction. Moreover, the density fields of the different species are all proportional to the mean profile $\Mstar(x)$, and the dispersion around this profile $\Mstar$ is Gaussian with variance $\sigma'/\sqrt{\log{N}}=\sigma$. Interestingly enough, this result holds even for $\alpha\ll \alpha_c$, i.e. far from the onset of patterns, as shown in Fig.~\ref{fig:profiles_and_dispersion_dmft_expansion}, where the amplitude of the patterns is of the same order as the mean density. 
Finally, although the system may display a time oscillating behavior and traveling waves, the prediction of such features is out of the scope of the present study, since they occur for larger $\sigma$ and for which the noise $\eta$ can no longer be simplified as it has been here.

\subsubsection{Case $\mu>0$: instability from the edge of the bulk}

In that case, at $\sigma=0$, the system splits into two families (for $d=1$) of periodic fields $M_a(x)$ and $M_b(x)$, of spatial periodicity $\lambda_p$, and  that satisfy 
\begin{align}
\begin{split}
0 =& D\nabla^2 M_a(x,t)\\ &+ M_a(x)K_\delta*\left[1-M_a(x) + \mu \frac{M_a(x)+M_b(x)}{2} \right]
\end{split}\\
\begin{split}
0 =& D\nabla^2 M_b(x,t)\\ &+ M_b(x)K_\delta*\left[1-M_b(x) + \mu \frac{M_a(x)+M_b(x)}{2} \right],
\end{split}
\end{align}
with, in addition, $M_a(x+\lambda_p/2)=M_b(x)$. The density profile, that is a random variable, can be written
\begin{align}
\rho^\star(x|Z_i) = Z_i M_a(x)+ (1-Z_i) M_b(x),
\end{align}
with $Z_i\in\{0,1\}$, a random variable that satisfies $\sum_i Z_i=N/2$ exactly, according to the PDE solutions. The $Z_i$ are thus correlated but the correlation vanishes in the $N\to \infty$ limit.

\begin{figure}
    \centering
\includegraphics[width=0.95\linewidth]{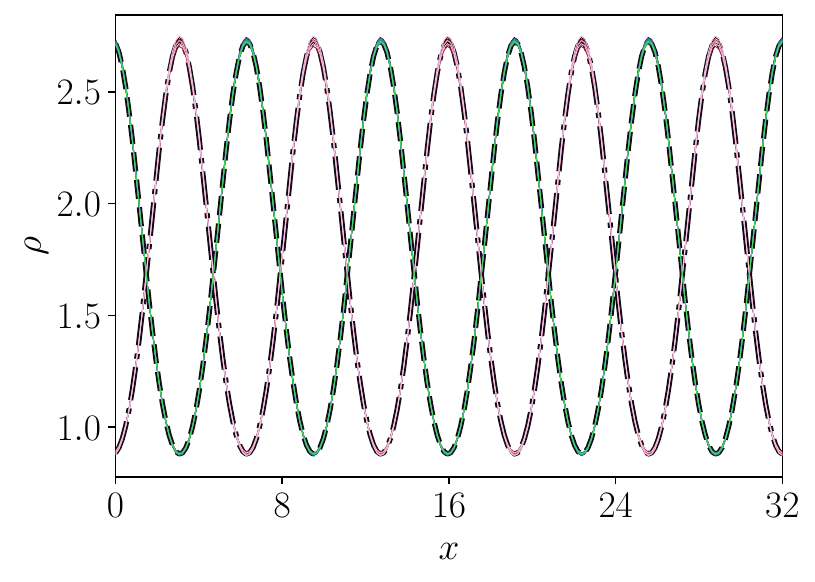}
    \caption{Intertwined fields $M_a(x)$ (black dashed line) and $M_b(x)$ (dotted-dashed line),  and a few density profiles (color lines) distributed around them.  Parameters: $\sigma=0.001$, $\mu=0.4$, $\gamma=0$, $\delta=4$, $D=0.30$, $N=500$, system size $L=32$, number of gridpoints $L_\mathrm{grid}=200$. The number of species closely distributed around $M_a$ is exactly $N/2$. We checked that identical results are obtained for other values of $\gamma$, which confirms that $\gamma$ does not play a role to leading order in $\sigma$.}
    \label{fig:profiles_and_dispersion_mu_positive}
\end{figure}
We now look at a small $\sigma $ expansion. The density is expanded as $\rho(x,t)=\rho^\star(x)+\sigma\psi(x,t)$.
The constraints on the noise correlation reads
\begin{align}
\langle \eta(x,t) \eta(x',t') \rangle &=  \langle \rho(x,t) \rho(x',t') \rangle\\
&= \frac{M_a(x)M_a(x')+M_b(x)M_b(x')}{2}+O(\sigma).
\end{align}
Since the noise is Gaussian, the covariance matrix of rank 2 fully determines the noise, which is then given by 
\begin{align}
    \eta(x) =\frac{\xi_a M_a(x)+\xi_b M_b(x)}{\sqrt{2}},
\end{align}
with $\xi_a$, $\xi_b$ independent Gaussian random variables of  mean 0 and variance 1. One can show that the stationary solution for $\psi(x)$ is not a simple linear combination of the two fields $M_a$ and $M_b$, but rather involves nonlocal operators. As such, the stationary solution cannot be obtained explicitly.  It has nonetheless the same flavor as the $\mu<0$ case, where each species follows one of the two master densities $M_a$ of $M_b$, as displayed in Fig.~\ref{fig:profiles_and_dispersion_mu_positive}. In particular, the variance of the Gaussian distribution of abundances around each of the master densities $M_a(x)$ and $M_b(x)$ is now space dependent.

\subsection{Larger values of $\sigma$}

\begin{figure}
    \centering
    \includegraphics[width=1\linewidth]{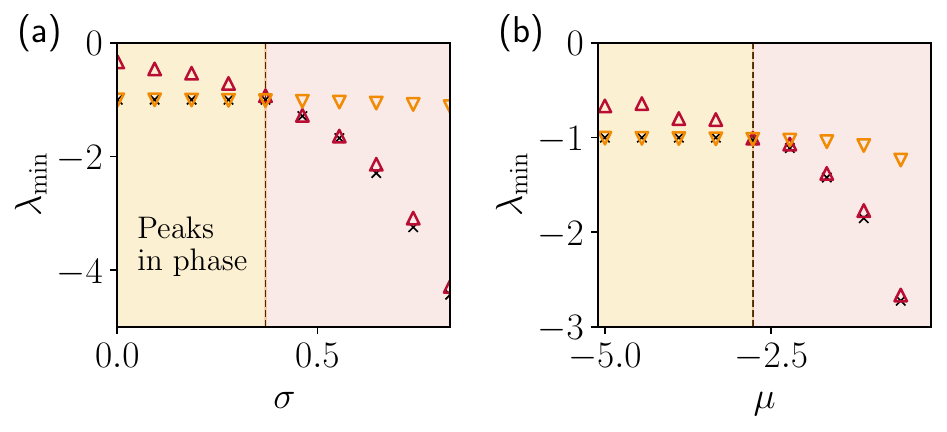}
    \caption{Comparison between true minimal eigenvalue ($\times$), the prediction $\lambda_\mathrm{min}^\mathrm{bulk}$ ($\vartriangle$), and the prediction of the outlier ($\triangledown$). Here, $N=1000$, $\gamma=0.7$, $\mu=-2$ in (a) and $\sigma=0.5$ in (b). Taking $\alpha$ such that patterns develop at the onset of instability, we observe density peaks in phase in the yellow shaded area, and dispersion of the phase in the red shaded area.}
    \label{fig:lambda_vs_sigma_mu_N1000}
\end{figure}

As the heterogeneity increases, mean cooperation or competition are lost in the noise of interactions with other species. As long as the instability at finite wavelength comes from the outlier eigenvalue of $RW$, species' patterns are in phase close to the instability onset. Their amplitudes can be strongly spread however. When the outlier eigenvalue is absorbed by the bulk, patterns no longer share the same phase. The transition between the two regimes is given in Appendix~\ref{app:rmt_results}, and is summarized in Fig.~\ref{fig:lambda_vs_sigma_mu_N1000} where the prediction of the outlier and the bulk minimum eigenvalue are compared. 

\section{Conclusion}

Let us summarize what we have achieved in this paper. Starting from a spatially-extended ecosystem of generalized Lotka-Volterra type where interactions between species are random but possibly correlated, we have computed the criterion for the loss of stability of the spatially homogeneous ecosystem. The instability at finite wavelength cannot develop if the Fourier transform of the interaction kernel is positive for all Fourier modes. The stability criterion depends on the most abundant species and can be captured by the ratio $D/\delta^2$, that measures how far species diffuse in one generation with respect to the interspecies interaction range. If species move fast enough, the system remains homogeneous. As the homogeneous phase loses its stability, the stationary patterned profiles are solutions of a nonlocal F-KPP equation. In addition, we have identified a BBP transition at the linear stability level that translates in a transition in the stationary state, where nonlinear terms are relevant. Expanding in the heterogeneity level parameter close to the patterned state, we have shown that an explicit solution of the DMFT equation could still be obtained in the stationary regime.

From a theoretical point of view, our work suggests that combining approaches from active matter systems, theoretical ecology and disordered systems is a promising path to unveil salient features of the spatial organization of interacting species. In our work, we have mainly focused on the behavior at long times in a closed environment. The dynamics of spreading of the species in an unbounded domain remains to be explored. We have also discarded the noise coming from the spatial diffusion and the one coming from population dynamics, that should ultimately be considered to understand the invasion and the extinction of species. Experimentally, the dynamics of many interacting bacterial strains start to be scrutinized at the lab~\cite{monmeyran2021four,hu_gore2022,alejandro_martinez_interfacial_2023}, and new experiments with genetically engineered strains, in the spirit of~\cite{curatolo2020cooperative}, offer a way to test the mechanism we have pinpointed that lead to pattern formation. In natural habitats, experiments are difficult to carry out, but data of quality are progressively obtained by means of environmental DNA~\cite{bohmann2014environmental,aglieri2021environmental}. Accurate maps on the repartition of species offer a new way to test predictions based on the complex system approaches~\cite{zelnik_how_2024}.

\acknowledgements
We warmly thank A. Altieri, J.W. Baron, M. Benzaquen, J.-P. Bouchaud, and J. Garnier-Brun for interesting discussions and precious insights. R.Z. thanks T. Arnoulx de Pirey for illuminating discussions.
This research was conducted within the Econophysics \& Complex Systems Research Chair, under the aegis of the Fondation du Risque, the Fondation de l’\'Ecole polytechnique, the \'Ecole polytechnique and Capital Fund Management.

\appendix

\section{Generalization to non-feasible ecosystems}
\label{app:non_feasible}
In general, given the interaction matrix $W$, for large $N$ or for large $\sigma$, it is possible to show that species extinction necessary occurs~\cite{galla_dynamically_2018}. The system is then said to be non-feasible.  The results from the section~\ref{sec:criterion_for_instability} might differ in this situation. We address this question below.

We consider a system that has undergone species extinction and we want to know if the final homogeneous state is stable with respect to spatial perturbation of the species abundances. The perturbation affects all the species, surviving ones as well as extinct ones (this can be justified if one allows species to migrate into the considered domain). For clarity, we reorder the species indices, such that, by denoting $N_s(\leq N)$ the number of surviving species, we have $i\in\{1,\dots,N_s\}$ which refers to surviving species while $i>N_s$ will refer to extinct species. At the same time, we reorder the interaction matrix and we denote by $Y$ the new interaction matrix that can be cast with 4 blocks:
\begin{align}
    Y=\begin{pmatrix}
Y_{S\to S} & Y_{E\to S} \\
Y_{S\to E} &  Y_{E\to E}
\end{pmatrix},
\end{align}
with $Y_{S\to S}$ refers to the interactions of surviving species on other surviving species, $Y_{E\to E}$ the interactions of extinct species on extinct ones, and $Y_{S\to E}$ the interactions of surviving species on extinct ones. In particular, the blocks $Y_{S\to S}$ and $Y_{E\to E}$ are square matrices of size $N_s\times N_s$ and $(N-N_s)\times (N-N_s)$, respectively. 
Considering a perturbation $\bm \psi(x,t)=(\psi_1,\dots,\psi_N)^\top$, the linearized dynamics of Eq.~\eqref{eq:model_SEGLV} reads
\begin{align}
\begin{split}
    \partial_t \hat\psi_i =& -Dk^2\hat\psi_i 
    + n_i\hat K_\delta(k)\sum_{j=1}^N Y_{ij}\hat\psi_j + (1+\sum_{j=1}^N Y_{ij}n_j)\hat\psi_i,
\end{split}
\end{align}
where now one has to be careful of the fact that for $i>N_s$, $n_i=0$. In a compact form, one can write
\begin{align}
    \partial_t \hat {\bm \psi} = (-Dk^2 I +U)\hat {\bm\psi},
\end{align}
with the matrix $U$ given by
\begin{align}
    U=\begin{pmatrix}
 \hat K_\delta(k) R_s Y_{S\to S} & \hat K_\delta(k) R_s Y_{E\to S}\\
0 & \mathrm{Diag}( 1+\sum_{j=1}^{N_s}Y_{ij}n_j)_{i\,\mathrm{extinct}}
    \end{pmatrix},
\end{align}
and $R_s$ is the diagonal matrix of size $N_s\times N_s$ with elements $n_i>0$, the positive abundances of the surviving species. The stability is then determined by the spectrum of the operator $-Dk^2 I +U$. In particular, since $U$ is block upper triangular, its spectrum is the union of the spectrum of  $\hat K_\delta(k) R_s Y_{S\to S}$ and the spectrum of $D_E\equiv \mathrm{Diag}(1+\sum_{j=1}^{N_s}Y_{ij}n_j)$. It turns out that the eigenvalues of $D_E$ are all necessarily negative since the extinct species have all died out dynamically in the homogeneous system by construction. The destabilization of the ecosystem can only come from the spectrum of  $\hat K_\delta(k) R_s Y_{S\to S}$.
The work~\cite{baron_breakdown_2023} has addressed the question of the boundaries of the spectrum of $R_s Y_{S\to S}$. Let us then discuss the stability depending on the sign of the kernel $\hat K_\delta$:

(i) If $\hat K_\delta(k)\geq 0$ then $\lambda_\mathrm{max}[-Dk^2 I +U] = -Dk^2+\hat K_\delta(k)\lambda_\mathrm{max}[ R_s Y_{S\to S}]$. The matrix $J'\equiv R_s Y_{S\to S}$ is referred to as the reduced Jacobian in Ref.~\cite{baron_breakdown_2023}. It turns out that the leading eigenvalue (which is real) of the reduced Jacobian $J'$ becomes positive exactly when the leading eigenvalue of the reduced interaction matrix $Y_{S\to S}$ becomes positive. However, as soon as $\lambdaMax(Y_{S\to S})>0$ the mode $k=0$ diverges. In other words, the reduced interaction matrix fully determines the stability in this case, and the spatially-extended system loses stability for the modes $k\to 0$ exactly when the zero-dimensional system loses stability. In this case we should not expect stable regular spatial heterogeneities. 

(ii) If for some $\tilde k$, one has $\hat K_\delta(\tilde k)\leq 0$, then $\lambda_\mathrm{max}[-D\tilde k^2 I +U] = -D{\tilde k}^2+\hat K_\delta(\tilde k)\lambda_\mathrm{min}[ R_s Y_{S\to S}]$.
Using perturbation theory like in the feasible ecosystem case, we find that one has
\begin{align}
    \lambda_\mathrm{max}[-Dk^2 I +U] = -D{\tilde k}^2+\hat K_\delta(\tilde k)\max_i n_i.
\end{align}
Again, the stability is set by the most abundant species.

\section{Known results in the zero-dimensional case}
\label{app:zero_dim_results}
 
We recall here the known results on the distribution of fixed points and the system stability when $d=0$, as it can be found in \cite{opper_phase_1992,galla_step_by_step_2024}.
We consider a generalized Lotka-Volterra dynamics in zero dimension
\begin{align}
\dot n_i = n_i(t) \left( 1-n_i(t)+\sum_{i=1}^N A_{ij}n_j(t) + h_i(t) \right),
\end{align}
where a perturbation field $h_i(t)$ has been introduced for convenience to compute the response to a perturbation on the dynamics. Via the cavity method (sketched in the next section) or via the generating functional analysis, one obtains the DMFT equation. This equation pinpoints the role of 3 observables that self-consistently determine the dynamics. We define the mean abundance of the species 
\begin{align}
    M(t)&\equiv \lim_{N\to\infty}\frac{1}{N}\sum_{i=1}^N n_i(t),
\end{align}
the correlation of the species abundance,
\begin{align}
    C(t,t')&\equiv \lim_{N\to \infty} \frac{1}{N}\sum_{i=1}^N n_i(t)n_i(t'),
    \end{align}
and the response function
\begin{align}
     G(t,t')&\equiv \lim_{N\to \infty} \frac{1}{N}\sum_{i=1}^N \frac{ \delta n_i(t)}{\delta h_i(t')}.
\end{align}

Assuming that we lie in the region of parameters where the system has converged to a stationary state, the final abundance of a species is a random variable that we denote $\nstar$. The first and the second moment of the $\nstar$ distribution are then denoted $M=\langle\nstar\rangle_\star$, $q= \langle\nstar^2\rangle_\star$, and the static response function is denoted $\chi\equiv \int_0^\infty G(\tau)\dd\tau$. Defining $Dz = \frac{dz}{\sqrt{2\pi}}e^{-z^2/2}$, one obtains $\chi$, $q$ and $M$ as the solution of the self-consistent equations:
\begin{align}
    \chi &= \frac{1}{1-\gamma \sigma^2 \chi}\int_{-\infty}^\Delta Dz\\
    M &= \frac{\sigma \sqrt{q}}{1-\gamma \sigma^2 \chi}\int_{-\infty}^\Delta (\Delta-z) Dz\\
    1 &= \frac{\sigma^2}{(1-\gamma\sigma^2\chi)^2}\int_{-\infty}^\Delta (\Delta-z)^2 Dz,
\end{align}
and $\Delta = \frac{1+\mu M}{\sigma\sqrt{q}}$.
In Ref.~\cite{galla_dynamically_2018}, the distribution of fixed points is computed and reads
\begin{align}
\nstar = \frac{1+\mu M+\sigma \sqrt{q}z}{1-\gamma\sigma^2\chi}\Theta\left( \frac{1+\mu M+\sigma \sqrt{q}z}{1-\gamma\sigma^2\chi}\right),
\label{eq:truncated_gaussian}
\end{align}
with $\Theta$ the Heaviside function and $z$ a Gaussian random variable of mean $0$ and variance $1$. The complete distribution of $\nstar$ thus splits between the surviving species (of fraction $\phi$) with distribution $p_+(\nstar)$, and the extinct species of fraction $1-\phi$.

\section{Dynamical-mean-field theory equations}
\label{app:dmft_theory}

In the main text, the phase diagram has been obtained using perturbation analysis, valid only at order 1 in $\sigma$. We have also used results from dynamical-mean field theory obtained in zero dimension~\cite{galla_dynamically_2018}. In what follows we will generalize the DMFT equation in dimension $d\geq1$ and see what one can deduce from this set of self-consistent equations. 

\subsection{Sketch of the derivation for the DMFT equations}

We derive the DMFT equations via the cavity method~\cite{mezard1987spin}. A new species in the system is considered as a perturbation of existing dynamics. Interactions between species are then interpreted as coming from a fluctuating bath. A set of an infinite number of deterministic equations simplifies into a single stochastic equation describing the density of a typical representative species in our system.
We refer the reader to \cite{Roy_2019} where similar derivations are carried out in a pedestrian way.
Starting from our set of $N$ equations describing the evolution of $N$ species, 
\begin{align}
\begin{split}
    \partial_t \rho_i(x,t) =& D \nabla^2\rho_i(x,t)\\
    &+\rho_i(x,t)K_\delta*(1+\sum_{j=1}^NW_{ij}\rho_j(x,t)+h_i(x,t))
    \label{eq:system_N_species}
    \end{split}
\end{align}  
and we introduced an auxiliary field $h_i(x,t)$, later set to zero, that provides the response and correlation functions necessary to close the dynamics. 
We now follow the common steps of  the DMFT to obtain the self-consistent equation:\\
1. Let the $\rho_i(x,t)_{i=1,\dots,N}$ be solutions of the system \eqref{eq:system_N_species}.\\
2. Add a new species, labeled with index $0$, which follows a similar dynamics:
\begin{align}
\begin{split}
    \partial_t \rho_0(x,t)=& D \nabla^2\rho_0(x,t) \\
    &+\rho_0(x,t)K_\delta*(1+\sum_{j=0}^N W_{0j}\tilde \rho_j(x,t)),
    \label{eq:rho_0_perturbation}
    \end{split}
\end{align}
where the $\tilde \rho_j(x,t)$ are the solutions of the dynamics that includes interactions with species $0$.\\
3. For $N$ large enough, the effect of a new species on the dynamics is small and can thus be treated via linear perturbation theory. The perturbed trajectories are thus given by
\begin{align}
    \tilde \rho_j(x,t)=\rho_j(x,t)+ \sum_{i=1}^N \int \dd x' \dd t' \chi_{ji}(x,x',t,t') W_{i0}\rho_0(x',t'),
    \label{eq:perturbed_rho_i}
\end{align}
where $\chi_{ji}(x,x',t,t')\equiv \frac{\delta \rho_j(x,t)}{\delta h_i(x',t')}$ denotes the response function.\\
4. We can insert Eq.~\eqref{eq:perturbed_rho_i} into Eq.~\eqref{eq:rho_0_perturbation} to obtain 
\begin{align}
\begin{split}
    \partial_t \rho_0(x,t)=&D\nabla^2 \rho_0(x,t)\\
    &+\rho_0(x,t)K_\delta*[1 + A_1(x,t) + A_2(x,t)]
\end{split}
\end{align}
with
\begin{align}
A_1(x,t) &=  \sum_{j=0}^N W_{0j}\rho_j(x,t) \\
A_2(x,t) &= \sum_{j=1}^N \sum_{i=1}^N W_{0j}\int \dd x' \dd t' \chi_{ji}(x,x',t,t')W_{i0}\rho_0(x',t').
\end{align}
5. This latter equation can simplify when using the correlations between the matrix coefficients of $W$, invoking the central limit theorem and neglecting terms scaling in $O(N^{-1/2})$, see \cite{galla_step_by_step_2024}.\\ 
6. We finally obtain the evolution equation of the density $\rho_0$, but since all species play a similar role, $\rho_0$ encodes in fact the ``typical'' behavior any species in the system. We thus drop the index $0$ and the evolution equation reads
\begin{widetext}
\begin{align}
        \partial_t \rho(x,t)= D\nabla^2 \rho(x,t) + \rho(x,t)K_\delta*\left(1-\rho(x,t) + \mu M(x,t)+\sigma \eta(x,t)+\gamma \sigma^2\int \dd x' \dd t' \chi(x,x',t,t')\rho(x',t') \right),
        \label{eq:dmft_equation_full}
\end{align}
\end{widetext}
where  $M(x,t) = \langle \rho(x,t) \rangle_\star$, $\eta(x,t)$ is a Gaussian noise with space and time correlation 
\begin{align}
    \langle \eta(x,t) \eta(x',t')\rangle_\star = \left\langle \rho(x,t) \rho(x',t')\right\rangle_\star
\end{align}
and the response function
\begin{align}
    \chi(x,x',t,t')&= \frac{1}{\sigma}\left\langle \frac{\delta \rho(x,t)}{\delta \eta(x',t')}\right\rangle_\star.
\end{align}
The average $\langle \cdot \rangle_\star$ is taken over several realizations of the complete dynamics \eqref{eq:model_SEGLV}.

\subsection{Fixed points and stability analysis}

There are a priori two kinds of fixed points in the DMFT: homogeneous fixed points, denoted $\nstar$ and non-homogeneous ones, $\rho^\star(x)$. Looking for homogeneous fixed points is equivalent to finding the ones from the zero-dimensional model, already presented in Appendix~\ref{app:zero_dim_results}.

On the other hand, we can find necessary conditions to the existence of a spatially dependent stationary state $\rho(x)$. Such a state would indeed solve the time-independent version of Eq.~\eqref{eq:dmft_equation_full}, and one solution has been obtained for small $\sigma$ and $\mu<0$. Determining the stability of such solutions via a DMFT linear stability analysis cannot be performed easily because of products of space-dependent fields that hinder modes diagonalization via Fourier transforms. Nonetheless, we suggest to follow a general path to assess state stability and will refine hypotheses later.

A natural procedure to assess the stability is to perturb the system with a multiplicative Gaussian white noise $\epsilon \rho(x,t)\xi(x,t)$ (with $\epsilon \to 0$) such that the perturbed dynamics still satisfies $\rho(x,t)>0$. One then computes the power spectrum of the field that is continuously kicked away from the fixed points by the excitations. As we expect to stay close to the fixed point solution, we write $\rho$ and $\eta$ as
\begin{align}
    \rho(x,t) &= \rho^\star(x) + \psi(x,t)\\
    \eta(x,t) &= \eta^\star(x) + \nu(x,t),
\end{align}
with $\|\psi\|,\, \|\nu\| \sim O(\epsilon)$, $\langle \psi(x,t) \rangle_{\xi,\nu}=0$,  $\langle \nu(x,t) \rangle_{\xi,\nu}=0$, $\langle \xi(x,t) \rangle_{\xi}=0$, and the self-consistent correlation relations 
\begin{align}
    \langle \rho^\star(x) \rho^\star(x')\rangle_{\star} &=  \langle \eta^\star(x)\eta^\star(x')\rangle_\star\\
    \langle \psi(x,t) \psi(x',t')\rangle_{\xi,\nu} &= \langle \nu(x,t)\nu(x',t')\rangle_{\xi,\nu},
    \label{eq:correlations_psi_nu_dmft}
\end{align}
where averaging over the white noise $\xi$, the DMFT noise $\nu$ and the distribution of fixed points $\rho^\star$ is indicated by the indices of the brackets. A step-by-step treatment of the different terms can be found in the lecture notes of Galla, see \cite{galla_step_by_step_2024}. 
Retaining leading order terms in Eq.~\eqref{eq:dmft_equation_full} leads to linear equations for $\psi(x,t)$. In particular, close to the identically zero fixed points $\rho^\star(x)\equiv 0$, one obtains
\begin{align}
    \partial_t\psi (x,t) =&  D \nabla^2 \psi(x,t)+ \psi(x,t) (1+\mu M+\sigma\sqrt{q}\eta^\star).
\end{align}
We note first that the perturbations around the identically (flat) zero fixed point vanish. Indeed in Fourier space the relaxation rate of a Fourier mode reads $- D k^2 + (1+\mu M +\sigma \sqrt{q}\eta^\star)$, with $(1+\mu M +\sigma \sqrt{q}\eta^\star)<0$ since $\rho^\star=0$ from the truncated Gaussian distribution in \eqref{eq:truncated_gaussian}. 
On the other hand, the fluctuations around a non-zero fixed point satisfy
\begin{align}
\begin{split}
    \partial_t\psi(x,t) =&  D \nabla^2 \psi(x,t)\\
    &+\rho^\star(x) K_\delta*\Big(-\psi(x,t) + \sigma\nu(x,t)+\epsilon \xi(x,t)\\
    &+\gamma \sigma^2\int \dd t' \dd x' \chi(x,x',t,t')\psi(x',t')\Big).
    \end{split}
    \label{eq:dynamics_fluctutation_DMFT}
\end{align}
A comprehensive study of this equation with a space dependent fixed point $\rho^\star(x)$ would be out of the scope of the present work. In the following we will focus on the perturbations around homogeneous nonzero fixed points.
Close to a homogeneous state of density $\nstar$, the dynamics \eqref{eq:dynamics_fluctutation_DMFT} in Fourier space reads
\begin{align}
\begin{split}
    i\omega \hat\psi(k,\omega)=&- D k^2 \hat \psi(k,\omega)+\nstar \hat K_\delta(k)\\
    &\times\Big(-\hat\psi(k,\omega) + \sigma\hat \nu(k,\omega)+\epsilon \hat\xi(k,\omega)\\
    &\quad +\gamma \sigma^2 \hat \chi(k,\omega) \hat\psi(k,\omega)\Big),
    \end{split}
\label{eq:fluctutation_DMFT_fourier}
\end{align}
which, after factorizing, yields
\begin{align}
\begin{split}
   \hat\psi(k,\omega)
    =\frac{ \sigma\hat \nu(k,\omega)
    +\epsilon \hat\xi(k,\omega)}{\frac{i\omega + D k^2}{\nstar \hat K_\delta(k)}+ 1-\gamma \sigma^2 \hat \chi(k,\omega)}.
    \label{eq:psi_k_omega}
    \end{split}
\end{align}
The power spectrum is obtained by taking the modulus squared and the average over the independent noises $\xi$, DMFT noise $\nu$, and realizations of fixed point $\nstar$.
In particular, the contributions to the spectrum from the $\nstar=0$ vanishes, which explicitly writes
\begin{align}
\begin{split}
    \langle|\psi(k,\omega)|^2\rangle_{\xi,\nu,\star}=& \int \dd\nstar [p_+(\nstar)+(1-\phi)\delta(\nstar)]\\
    &\times \langle|\psi(k,\omega)|^2\rangle_{\xi,\nu} 
    \end{split}\\
    =&\int \dd\nstar p_+(\nstar)\langle|\psi(k,\omega)|^2\rangle_{\xi,\nu} \\
    \equiv& \langle|\psi(k,\omega)|^2\rangle_{\xi,\nu,\star+}.
\end{align}
Using now the independence of $\xi$, $\nu$ and $\nstar$, and using Eq.~\eqref{eq:correlations_psi_nu_dmft}, one obtains
\begin{align}
    \langle |\hat \psi(k,\omega)|^2 \rangle_{\xi,\nu,\star+} = \frac{\epsilon^2H(k,\omega)}{1-\sigma^2 H(k,\omega)},
\end{align}
with 
\begin{align}
    H(k,\omega)= \left\langle \left|\frac{i\omega+D k^2}{\nstar \hat K_\delta(k)}+1-\gamma \sigma^2\hat \chi(k,\omega) \right|^{-2} \right\rangle_{\star+},
\end{align}
and we check that for $\hat K_\delta(k)=0$, the power spectrum remains finite. The solution to $\langle |\hat \psi(k,\omega)|^2 \rangle_{\xi,\nu,\star+}\to\infty$ reads
\begin{align}
    H(k,\omega) =\frac{1}{\sigma^2}.
    \label{eq:diverging_spectrum_k}
\end{align}

We now take the $\omega\to 0$ limit to probe the behavior of the spectrum at long times, and we will denote $H(k)\equiv H(k,\omega=0)$, and $\hat\chi(k)\equiv \hat \chi(k,\omega=0)$, which is real. We check that taking the $k\to 0$ limit yields back the power spectrum of the zero-dimensional case investigated in \cite{opper_phase_1992,bunin_ecological_2017,galla_dynamically_2018}, that reads
\begin{align}
    \langle |\hat \psi(0,0)|^2 \rangle_{\xi,\nu,\star+} = \frac{\epsilon^2 \phi}{  [1-\gamma \sigma^2 \hat \chi(0)]^2- \phi\sigma^2},
\end{align}
and which diverges when one has $\phi\sigma^2=(1-\gamma \sigma^2 \chi)^2$, with $\chi = \hat \chi(k=0)$.
Keeping the mode $k$ dependence, the onset of instability of the system is the manifold on which the power spectrum starts diverging. The condition $\frac{\dd  \langle |\hat \psi(k,\omega)|^2 \rangle_{\xi,\nu,\star+}}{\dd k}|_{k_c} =0$ yields the fastest growing mode and reads
\begin{align}
    \epsilon^2\frac{H'(k_c)}{[1-\sigma^2 H(k_c)]^2} =0,
    \label{eq:H_prime_kc}
\end{align}
with 
\begin{align}
\begin{split}
    H'(k)=-2\left\langle\frac{\frac{2Dk}{n^\star \hat K_\delta(k)}-\frac{Dk^2 \hat K'_\delta(k)}{n^\star\hat K_\delta^2(k)}-\gamma\sigma^2\hat\chi'(k)}{J(n^\star,k)^3} \right\rangle_{\star+},
    \label{eq:maximum_k_dmft}
    \end{split}
\end{align}
and where we defined 
\begin{align}
    J(n^\star,k) = \frac{D k^2}{n^\star \hat K_\delta(k)}+1-\gamma\sigma^2\hat\chi(k).
\end{align}
Equation~\eqref{eq:maximum_k_dmft} can be further simplified. Indeed, through Eq.~\eqref{eq:psi_k_omega}, we have, by definition of the response function,
\begin{align}
    \hat \chi(k,\omega)&=\left\langle \frac{\partial \psi(k,\omega)}{\partial [\epsilon \hat \xi(k,\omega)]} \right\rangle_{\star}\\
    &= \int_0^\infty \frac{[p_+(n^\star) +(1-\phi)\delta(n^\star)]\dd n^\star}{\frac{i\omega + D k^2}{\nstar \hat K_\delta(k)}+ 1-\gamma \sigma^2 \hat \chi(k,\omega)}.
    \label{eq:chi_k_omega_self_consistent}
\end{align}
By taking the derivative with respect to $k$ in Eq.~\eqref{eq:chi_k_omega_self_consistent} (and taking $\omega\to0$), one obtains
\begin{align}
\begin{split}
    \hat\chi'(k) =& \left(\frac{2Dk}{ \hat K_\delta(k)}-\frac{Dk^2 \hat K'_\delta(k)}{\hat K_\delta^2(k)}\right)\left\langle\frac{1}{n^\star J(n^\star,k)^2}\right\rangle_{\star}\\
    &\times \frac{1}{1+\left\langle\frac{\gamma\sigma^2}{J(n^\star,k)^2}\right\rangle_{\star}}.
    \end{split}
\end{align}
Injecting this expression in Eq.~\eqref{eq:maximum_k_dmft}, we find that $H'(k_c)=0$ only if $k_c$ solves $2=k_c \hat K_\delta'(k_c)/\hat K_\delta(k_c)$. We thus recover here the condition of Eq.~\eqref{eq:definition_k_c} on the critical mode that we obtained from the linear stability analysis conducted in Section~\ref{sec:criterion_for_instability}. Finally using a mean-field approximation in Eq.~\eqref{eq:chi_k_omega_self_consistent}, we find that $\hat \chi(k)$ is a root of a second-degree polynomial, which leads to
\begin{align}
    \hat \chi(k) = \frac{1+\frac{D k^2}{M \hat K_\delta(k)}+\sqrt{\left(1+\frac{D k^2}{M \hat K_\delta(k)}\right)^2-4\gamma\sigma^2}}{2\gamma \sigma^2}.
    \label{eq:chi_explicit_dmft}
\end{align}
This expression can then be injected into $H(k)$ in order to solve the instability criterion $H(k)=1/\sigma^2$.

 We now examine the case where $\hatKdelta$ can be negative and we focus on the fastest growing mode $k=k_c$, for which $\hatKdelta(k_c)<0$ necessarily. One would like to assess the expectation $H(k_c)$, but one notices that there is a pole $n^\star_\mathrm{pole}(k)=-
\frac{D k^2}{ \hat K_\delta(k) [1-\gamma \sigma^2\hat \chi(k)]}$ in the integrand of $H(k)$ that leads to divergence of $H(k)$ as soon as the support of the distribution of $\nstar$ intercepts it.  In addition, since $H(k_0)= 0$ for modes $k_0$ that satisfy $\hatKdelta(k_0)=0$, the function $k\mapsto H(k)$ is surjective on $\mathbb{R}_+$, hence there exists a mode $k$ such that Eq.~\eqref{eq:diverging_spectrum_k} is satisfied, yielding the divergence of the power spectrum for that mode. In other words, the system is always unstable when the number of species $N$ goes to infinity, in line with our findings of Section~\ref{sec:patterns_and_phase_diag}.

To avoid the trivial divergence of the dynamics, we finally consider a large but finite system $(1\ll N<\infty)$, and it is interesting to restrict the integration bound on $\nstar$ to the highest abundance that can be obtained from $N$ draws on the fixed point distribution. Denoting by $n_\mathrm{max}^\star=\max_{1\leq i\leq N}n_i^\star$, we define
\begin{align}
     H_N(k,n_\mathrm{max}^\star) = \int_0^{n_\mathrm{max}^\star} \frac{\dd n^\star p_+(n^\star) }{ \left(\frac{D k^2}{\nstar \hat K_\delta(k)}+ 1-\gamma \sigma^2\hat \chi(k)\right)^{2}}.
     \label{eq:H_N_integral}
\end{align}
If $n_\mathrm{max}^\star<\inf_{k}n^\star_\mathrm{pole}(k)$, then $H_N(k,n_\mathrm{max}^\star)$ remains finite and is maximal for $k=k_c$, see Eq.~\eqref{eq:H_prime_kc}. Hence we have
\begin{align}
    H_N(k,n_\mathrm{max}^\star)\leq H_N(k_c,n_\mathrm{max}^\star),
\end{align}
and the power spectrum $\langle|\psi(k,\omega)|^2\rangle_{\xi,\nu,\star}$ diverges as soon as 
\begin{align}
    \frac{1}{\sigma^2}= H_N(k_c,n_\mathrm{max}^\star).
\end{align}
This equation can be solved numerically to recover the transition line in parameter space $(\alpha,\sigma,\mu,\gamma,N)$, see Fig.~\ref{fig:transition_lines}. The agreement between the DMFT and the perturbation theory is remarkable.

\section{Exact results from random matrix theory}
\label{app:rmt_results}

In the presence of the interaction kernel $K_\delta$, we have seen that the stability of the ecosystem is determined by the minimal eigenvalue of the matrix $RW=R(-I+A)$. We are computing this minimal eigenvalue in what follows. 

We first write the matrix in the following way,
\begin{equation}
    RW = -R + \frac{\sigma}{\sqrt{N}} RZ -\frac{\mu}{N}R + \frac{\mu}{N} R u u^\top,
\end{equation}
with $(Z)_{ij} =z_{ij}$ and $u u^\top$ the matrix filled of $1$, and such that it is clear that the effect of $\mu$ is that of a rank one perturbation in the $N\to\infty$ limit.

Consider first the case $\sigma=0$. The species are all equivalent, and $R=\diag(M,\dots, M)$, with $M=1/(1-\mu) +O(1/N)$. The spectrum $RW$ can be computed analytically, and the eigenvalues are simply 
\begin{align}
    \lambda(RW)= \{-1 ,-M,\dots,-M\}+ O(1/N).
\end{align}
This means that asymptotically we have $N-1$ eigenvalues equal to $-M$ and one eigenvalue equal to $-1$. Therefore, the minimum depends on the sign of $\mu$,
\begin{equation}
    \lambda_{\min} = \left\lbrace\begin{array}{cc}
     -1    & \textrm{ for } \mu < 0\\
     \frac{1}{\mu - 1}   &  \textrm{ for } \mu > 0.
    \end{array}\right.
\end{equation}

The general case will behave in a similar way. It is useful to first consider the case $\sigma>0$, and $\mu=0$. We will generalize to the case $\mu\neq 0$ later. 

We introduce the resolvent for the matrix $RW$, 
\begin{equation}
    G(z) = (zI - RW)^{-1},
\end{equation}
and its trace,
\begin{equation}
    g(z) = \frac{1}{N} \textrm{Tr}\,G(z) = \frac{1}{N} \sum_{i=1}^N G_{ii}(z)
\end{equation}
The function $g(z)$ encodes the information about the eigenvalues of $RW$ in its poles. We will approximate it by means of the cavity method. Using the Schur complement formula, we have
\begin{align}
    G_{ii}(z) = \frac{1}{z + n_i -n_i \sigma^2 \frac{1}{N}\sum_{jk} z_{ij}z_{ki}n_k G_{jk}^{(i)}(z)},
    \label{eq:Schur-resolvent}
\end{align}
where $G^{(i)}$ is the resolvent for the matrix $RW$ with row and column $i$ removed. Using the Schur complement a second time we can write an equations for it as well,
\begin{align}
    G_{jj}^{(i)}(z) = \frac{1}{z + n_j - n_j\sigma^2 \frac{1}{N} \sum_{k\ell} z_{jk} z_{\ell j}n_\ell G_{k\ell}^{(i,j)}(z)}
    \label{eq:Schur-cavity}
\end{align}
In the large $N$ limit we can use the statistical properties of the $z_{ij}$ variables and assume statistical properties of will be independent of the removed site, $G^{(i)}(z) \approx G^{(0)}(z)$, giving
\begin{equation}
    \frac{1}{N} \sum_{jk}z_{ij}z_{ki}n_k G^{(i)}_{jk}(z) \underset{N \gg 1}{\longrightarrow} \gamma g^{(0)}(z)
\end{equation}
where
\begin{equation}
    g^{(0)}(z) = \frac{1}{N} \sum_{j} n_j G^{(0)}_{jj}(z).
    \label{eq:g-cavity}
\end{equation}
Plugging this back in \eqref{eq:Schur-resolvent}, we get
\begin{equation}
    g(z) = \frac{1}{N}\sum_{i=1}^N \frac{1}{z + n_i - n_i \gamma \sigma^2 g^{(0)}(z)}.
\end{equation}
In order to write a self-consistent equation for $g^{(0)}(z)$ we assume statistical equivalence between $G^{(i)}$ and $G^{(i,j)}$, which should hold in the large $N$ limit, combining this with \eqref{eq:Schur-cavity} and \eqref{eq:g-cavity} we write down the following equation,
\begin{equation}
    g^{(0)}(z) = \frac{1}{N}\sum_{i=1}^N \frac{n_i}{z + n_i - n_i \gamma \sigma^2 g^{(0)}(z)}.
    \label{eq:cavity}
\end{equation}
We can immediately see that for $0\le\sigma\ll 1$, $g(z)$ can be approximated with
\begin{equation}
    g(z) \approx \frac{1}{N}\sum_{i}\frac{1}{z + n_i},
\end{equation}
and the poles are all simply located at the values $-n_i$. Since we are interested in the minimum eigenvalue of $RW$, we can see it has to be associated with $n_{\max} = \max_i n_i$, giving $\lambda_{\min} = -n_{\max}$. 

\begin{figure}
    \centering
    \includegraphics[width=1\linewidth]{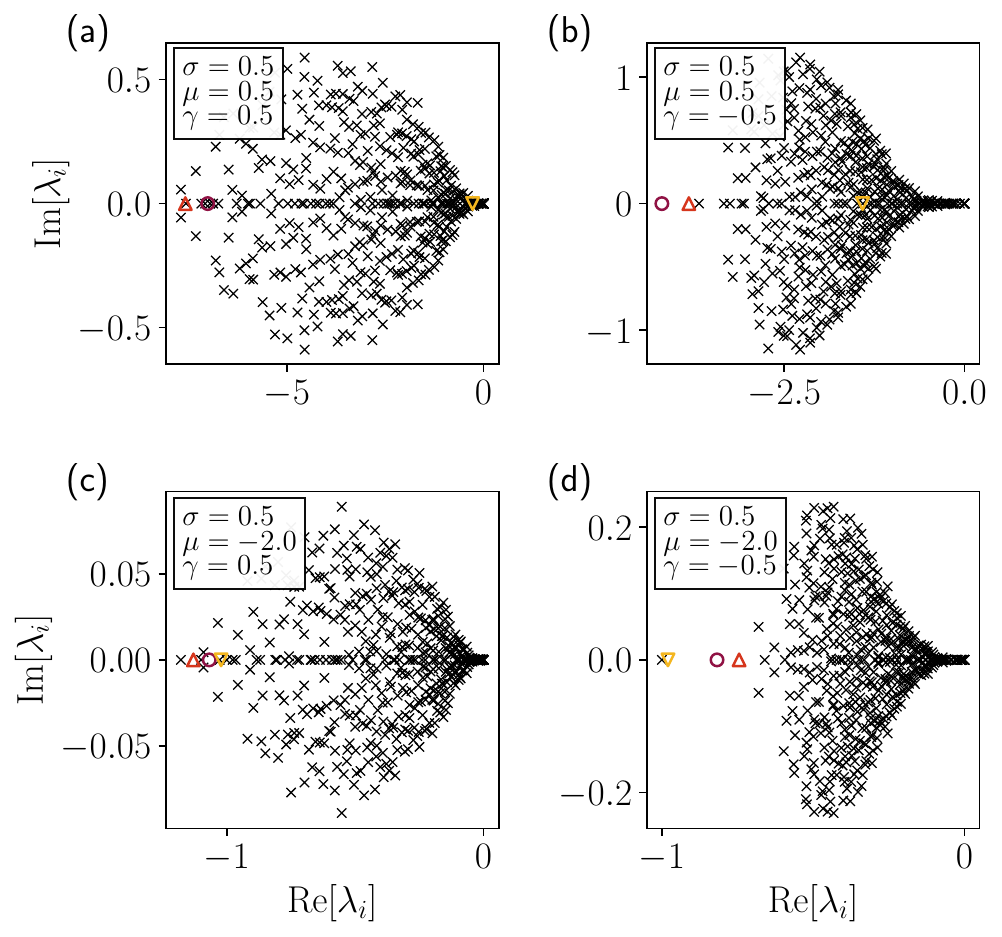}
    \caption{Eigenvalues in complex plane of the matrix $RW=R(-I+A)$ for different values of $(\sigma, \mu, \gamma)$, and predictions of the minimum eigenvalue with the different approaches. ($\circ$) naive perturbation theory yielding $\lambda_\mathrm{min}=-n_\mathrm{max}$, ($\vartriangle$) mean-field computation from the cavity method on the resolvent matrix, and ($\triangledown$) prediction of the outlier. Here $N=500$.}
    \label{fig:spectrum_from_RMT}
\end{figure}

Now assume that for moderate $\sigma$, the minimum eigenvalue is real and just slightly perturbed, $\lambda_{\min} = -n_{\max} - \Delta $, we can write down the equation for $\Delta$ assuming it is associated with a new pole, $g(-n_{\max} - \Delta) =\infty$,
\begin{equation}
    \Delta =  - n_{\max} \gamma \sigma^2 g^{(0)}(-n_{\max} - \Delta)
    \label{eq:pole}
\end{equation}
While \eqref{eq:cavity} and \eqref{eq:pole} constitute a well-defined system of equations for $g^{(0)}$ and $\Delta$, we can proceed further analytically if we perform a mean-field approximation for $g^{(0)}(z)$,
\begin{equation}
    g^{(0)}_\mathrm{mf}(z) = \frac{\bar{n}}{z + \bar{n} - \bar{n}\gamma \sigma^2 g^{(0)}_\mathrm{mf}(z)},
\end{equation}
where $\bar{n} = \frac{1}{N} \sum_i n_i$. This function has a simple analytical form,
\begin{equation}
    g^{(0)}_\mathrm{mf}(z) = \frac{z  + \bar{n} + \sqrt{(z + \bar{n})^2 - 4 \bar{n}^2 \gamma \sigma^2}}{2 \bar{n}\gamma \sigma^2}.
    \label{eq:cavity-MF}
\end{equation}
Note that we have precisely recovered the response function obtained in Eq.~\eqref{eq:chi_explicit_dmft}, if one sets $z=Dk^2/\hat K_\delta(k)$.
Using \eqref{eq:cavity-MF} with \eqref{eq:pole} we can derive
\begin{equation}
    \Delta = -\frac{n_{\max}}{2} + \frac{n_{\max}}{2}\sqrt{1+\frac{ 4 M \gamma \sigma^2}{n_{\max} - M} },
\end{equation}
where we have used the fact $\bar{n} = M$ for large $N$ by definition of $M$. All in all, the minimal eigenvalue reads
\begin{align}
    \lambda_{\min} &= -n_{\max} \left(\frac{1}{2}+\frac{1}{2}\sqrt{1+\frac{4M\gamma\sigma^2}{n_{\max}-M}} \right).
    \label{eq:eigenvalue_mini_no_mu}
\end{align}
Comparison with the order 0 perturbation theory is displayed in Fig.~\ref{fig:spectrum_from_RMT} where the full spectrum of the matrix $RW$ is computed for various $\sigma$, $\gamma$ and $\mu$. There, the matrix $R$ is obtained from running the 0-dimensional Lotka-Volterra dynamics given by $W$.  Hence, the matrices $R$ and $W$ are not independent, but assuming they are is indeed a very good approximation.

For the outlier we can follow similar steps as in \cite{baron2022eigenvalue}. The outlier eigenvalue should satisfy by definition $\det(\lambda_{\textrm{out}}I- RW) = 0$, which here writes
\begin{equation}
    \det[\lambda_{\textrm{out}}I- R(-I+\frac{\sigma}{\sqrt{N}}Z-\frac{\mu}{N}I) - \frac{\mu}{N} Ru u^{\top}]= 0
\end{equation}
If $\lambda_{\textrm{out}}$ is outside of the bulk then we should have $[\lambda_\textrm{out}I - R(-I+\frac{\sigma}{\sqrt{N}}Z-\frac{\mu}{N}I)]$ invertible, giving
\begin{equation}
    \det[I - \frac{\mu}{N}Ruu^{\top}G(\lambda_{\textrm{out}})] = 0
\end{equation}
using Sylvester's determinant theorem
\begin{equation}
    1 - \frac{\mu}{N}u^{\top}G(\lambda_{\textrm{out}})Ru = 0
\end{equation}
where by definition we have
\begin{equation}
    \frac{1}{N}u^{\top}G(\lambda_{\textrm{out}})Ru  = \frac{1}{N}\sum_{ij} n_{j} G_{ij}(\lambda_\textrm{out})
\end{equation}
If we approximate by keeping only the diagonal terms when $\sigma$ is small and use definition \eqref{eq:g-cavity}, we find the following equation for the outlier eigenvalue,
\begin{equation}
    1 - \mu g^{(0)}(\lambda_{\textrm{out}}) = 0
\end{equation}
Using the mean-field approximation, \eqref{eq:cavity-MF}, we find
\begin{equation}
    \lambda_{\textrm{out}} = -1  + \frac{\gamma \sigma^2}{\mu(1-\mu)}
\end{equation}
Note that this formula uses both a mean-field and a diagonal approximation, but the outlier is properly predicted for large values of $\mu$ and $\sigma$, as shown in Fig.~\ref{fig:lambda_vs_sigma_mu_N1000}. Also, the formula is no longer valid if $|\mu|\ll\sigma^2$. Finally equating the outlier eigenvalue and the minimum eigenvalue of the bulk obtained in Eq.~\eqref{eq:eigenvalue_mini_no_mu} yields the approximated frontier of the BBP transition.


\bibliography{active_Lotka_Volterra.bib}

\begin{thebibliography}{54}%
\makeatletter
\providecommand \@ifxundefined [1]{%
 \@ifx{#1\undefined}
}%
\providecommand \@ifnum [1]{%
 \ifnum #1\expandafter \@firstoftwo
 \else \expandafter \@secondoftwo
 \fi
}%
\providecommand \@ifx [1]{%
 \ifx #1\expandafter \@firstoftwo
 \else \expandafter \@secondoftwo
 \fi
}%
\providecommand \natexlab [1]{#1}%
\providecommand \enquote  [1]{``#1''}%
\providecommand \bibnamefont  [1]{#1}%
\providecommand \bibfnamefont [1]{#1}%
\providecommand \citenamefont [1]{#1}%
\providecommand \href@noop [0]{\@secondoftwo}%
\providecommand \href [0]{\begingroup \@sanitize@url \@href}%
\providecommand \@href[1]{\@@startlink{#1}\@@href}%
\providecommand \@@href[1]{\endgroup#1\@@endlink}%
\providecommand \@sanitize@url [0]{\catcode `\\12\catcode `\$12\catcode
  `\&12\catcode `\#12\catcode `\^12\catcode `\_12\catcode `\%12\relax}%
\providecommand \@@startlink[1]{}%
\providecommand \@@endlink[0]{}%
\providecommand \url  [0]{\begingroup\@sanitize@url \@url }%
\providecommand \@url [1]{\endgroup\@href {#1}{\urlprefix }}%
\providecommand \urlprefix  [0]{URL }%
\providecommand \Eprint [0]{\href }%
\providecommand \doibase [0]{https://doi.org/}%
\providecommand \selectlanguage [0]{\@gobble}%
\providecommand \bibinfo  [0]{\@secondoftwo}%
\providecommand \bibfield  [0]{\@secondoftwo}%
\providecommand \translation [1]{[#1]}%
\providecommand \BibitemOpen [0]{}%
\providecommand \bibitemStop [0]{}%
\providecommand \bibitemNoStop [0]{.\EOS\space}%
\providecommand \EOS [0]{\spacefactor3000\relax}%
\providecommand \BibitemShut  [1]{\csname bibitem#1\endcsname}%
\let\auto@bib@innerbib\@empty
\bibitem [{\citenamefont {Lotka}(1920)}]{lotka1920undamped}%
  \BibitemOpen
  \bibfield  {author} {\bibinfo {author} {\bibfnamefont {A.~J.}\ \bibnamefont
  {Lotka}},\ }\bibfield  {title} {\bibinfo {title} {Undamped oscillations
  derived from the law of mass action.},\ }\href@noop {} {\bibfield  {journal}
  {\bibinfo  {journal} {Journal of the american chemical society}\ }\textbf
  {\bibinfo {volume} {42}},\ \bibinfo {pages} {1595} (\bibinfo {year}
  {1920})}\BibitemShut {NoStop}%
\bibitem [{\citenamefont {Volterra}(1926)}]{volterra1926fluctuations}%
  \BibitemOpen
  \bibfield  {author} {\bibinfo {author} {\bibfnamefont {V.}~\bibnamefont
  {Volterra}},\ }\bibfield  {title} {\bibinfo {title} {Fluctuations in the
  abundance of a species considered mathematically},\ }\href@noop {} {\bibfield
   {journal} {\bibinfo  {journal} {Nature}\ }\textbf {\bibinfo {volume}
  {118}},\ \bibinfo {pages} {558} (\bibinfo {year} {1926})}\BibitemShut
  {NoStop}%
\bibitem [{\citenamefont {Maslov}\ and\ \citenamefont
  {Sneppen}(2017)}]{maslov2017population}%
  \BibitemOpen
  \bibfield  {author} {\bibinfo {author} {\bibfnamefont {S.}~\bibnamefont
  {Maslov}}\ and\ \bibinfo {author} {\bibfnamefont {K.}~\bibnamefont
  {Sneppen}},\ }\bibfield  {title} {\bibinfo {title} {Population cycles and
  species diversity in dynamic kill-the-winner model of microbial ecosystems},\
  }\href@noop {} {\bibfield  {journal} {\bibinfo  {journal} {Scientific
  reports}\ }\textbf {\bibinfo {volume} {7}},\ \bibinfo {pages} {1} (\bibinfo
  {year} {2017})}\BibitemShut {NoStop}%
\bibitem [{\citenamefont {Joseph}\ \emph {et~al.}(2020)\citenamefont {Joseph},
  \citenamefont {Shenhav}, \citenamefont {Xavier}, \citenamefont {Halperin},\
  and\ \citenamefont {Pe’er}}]{joseph2020compositional}%
  \BibitemOpen
  \bibfield  {author} {\bibinfo {author} {\bibfnamefont {T.~A.}\ \bibnamefont
  {Joseph}}, \bibinfo {author} {\bibfnamefont {L.}~\bibnamefont {Shenhav}},
  \bibinfo {author} {\bibfnamefont {J.~B.}\ \bibnamefont {Xavier}}, \bibinfo
  {author} {\bibfnamefont {E.}~\bibnamefont {Halperin}},\ and\ \bibinfo
  {author} {\bibfnamefont {I.}~\bibnamefont {Pe’er}},\ }\bibfield  {title}
  {\bibinfo {title} {Compositional lotka-volterra describes microbial dynamics
  in the simplex},\ }\href@noop {} {\bibfield  {journal} {\bibinfo  {journal}
  {PLoS computational biology}\ }\textbf {\bibinfo {volume} {16}},\ \bibinfo
  {pages} {e1007917} (\bibinfo {year} {2020})}\BibitemShut {NoStop}%
\bibitem [{\citenamefont {Dedrick}\ \emph {et~al.}(2023)\citenamefont
  {Dedrick}, \citenamefont {Warrier}, \citenamefont {Lemon},\ and\
  \citenamefont {Momeni}}]{dedrick2023does}%
  \BibitemOpen
  \bibfield  {author} {\bibinfo {author} {\bibfnamefont {S.}~\bibnamefont
  {Dedrick}}, \bibinfo {author} {\bibfnamefont {V.}~\bibnamefont {Warrier}},
  \bibinfo {author} {\bibfnamefont {K.~P.}\ \bibnamefont {Lemon}},\ and\
  \bibinfo {author} {\bibfnamefont {B.}~\bibnamefont {Momeni}},\ }\bibfield
  {title} {\bibinfo {title} {When does a {Lotka}-{Volterra} model represent
  microbial interactions? {Insights} from in vitro nasal bacterial
  communities},\ }\href {https://doi.org/10.1128/msystems.00757-22} {\bibfield
  {journal} {\bibinfo  {journal} {mSystems}\ }\textbf {\bibinfo {volume} {8}},\
  \bibinfo {pages} {e0075722} (\bibinfo {year} {2023})}\BibitemShut {NoStop}%
\bibitem [{\citenamefont {Fisher}(1937)}]{fisher_wave_1937}%
  \BibitemOpen
  \bibfield  {author} {\bibinfo {author} {\bibfnamefont {R.~A.}\ \bibnamefont
  {Fisher}},\ }\bibfield  {title} {\bibinfo {title} {The {Wave} of {Advance} of
  {Advantageous} {Genes}},\ }\href
  {https://doi.org/10.1111/j.1469-1809.1937.tb02153.x} {\bibfield  {journal}
  {\bibinfo  {journal} {Annals of Eugenics}\ }\textbf {\bibinfo {volume} {7}},\
  \bibinfo {pages} {355} (\bibinfo {year} {1937})}\BibitemShut {NoStop}%
\bibitem [{\citenamefont {Kolmogorov}\ \emph {et~al.}(1937)\citenamefont
  {Kolmogorov}, \citenamefont {Petrovsky},\ and\ \citenamefont
  {Piskunov}}]{kolmogorov1937}%
  \BibitemOpen
  \bibfield  {author} {\bibinfo {author} {\bibfnamefont {A.}~\bibnamefont
  {Kolmogorov}}, \bibinfo {author} {\bibfnamefont {I.}~\bibnamefont
  {Petrovsky}},\ and\ \bibinfo {author} {\bibfnamefont {N.}~\bibnamefont
  {Piskunov}},\ }\bibfield  {title} {\bibinfo {title} {Investigation of a
  diffusion equation connected to the growth of materials, and application to a
  problem in biology},\ }\href@noop {} {\bibfield  {journal} {\bibinfo
  {journal} {Bull. Univ. Moscow, Ser. Int. Sec. A}\ }\textbf {\bibinfo {volume}
  {1}} (\bibinfo {year} {1937})}\BibitemShut {NoStop}%
\bibitem [{\citenamefont {Reichenbach}\ \emph {et~al.}(2008)\citenamefont
  {Reichenbach}, \citenamefont {Mobilia},\ and\ \citenamefont
  {Frey}}]{reichenbach2008self}%
  \BibitemOpen
  \bibfield  {author} {\bibinfo {author} {\bibfnamefont {T.}~\bibnamefont
  {Reichenbach}}, \bibinfo {author} {\bibfnamefont {M.}~\bibnamefont
  {Mobilia}},\ and\ \bibinfo {author} {\bibfnamefont {E.}~\bibnamefont
  {Frey}},\ }\bibfield  {title} {\bibinfo {title} {Self-organization of mobile
  populations in cyclic competition},\ }\href@noop {} {\bibfield  {journal}
  {\bibinfo  {journal} {Journal of Theoretical Biology}\ }\textbf {\bibinfo
  {volume} {254}},\ \bibinfo {pages} {368} (\bibinfo {year}
  {2008})}\BibitemShut {NoStop}%
\bibitem [{\citenamefont {Täuber}(2024)}]{tauber2024}%
  \BibitemOpen
  \bibfield  {author} {\bibinfo {author} {\bibfnamefont {U.~C.}\ \bibnamefont
  {Täuber}},\ }\href {https://arxiv.org/abs/2405.05006} {\bibinfo {title}
  {Stochastic spatial lotka-volterra predator-prey models}} (\bibinfo {year}
  {2024}),\ \Eprint {https://arxiv.org/abs/2405.05006} {arXiv:2405.05006
  [cond-mat.stat-mech]} \BibitemShut {NoStop}%
\bibitem [{\citenamefont {Deforet}\ \emph {et~al.}(2019)\citenamefont
  {Deforet}, \citenamefont {Carmona-Fontaine}, \citenamefont {Korolev},\ and\
  \citenamefont {Xavier}}]{deforet2019}%
  \BibitemOpen
  \bibfield  {author} {\bibinfo {author} {\bibfnamefont {M.}~\bibnamefont
  {Deforet}}, \bibinfo {author} {\bibfnamefont {C.}~\bibnamefont
  {Carmona-Fontaine}}, \bibinfo {author} {\bibfnamefont {K.~S.}\ \bibnamefont
  {Korolev}},\ and\ \bibinfo {author} {\bibfnamefont {J.~B.}\ \bibnamefont
  {Xavier}},\ }\bibfield  {title} {\bibinfo {title} {Evolution at the edge of
  expanding populations},\ }\href {https://doi.org/10.1086/704594} {\bibfield
  {journal} {\bibinfo  {journal} {The American Naturalist}\ }\textbf {\bibinfo
  {volume} {194}},\ \bibinfo {pages} {291} (\bibinfo {year} {2019})},\ \bibinfo
  {note} {pMID: 31553215}\BibitemShut {NoStop}%
\bibitem [{\citenamefont {Lee}\ \emph {et~al.}(2022)\citenamefont {Lee},
  \citenamefont {Gore},\ and\ \citenamefont {Korolev}}]{lee2022}%
  \BibitemOpen
  \bibfield  {author} {\bibinfo {author} {\bibfnamefont {H.}~\bibnamefont
  {Lee}}, \bibinfo {author} {\bibfnamefont {J.}~\bibnamefont {Gore}},\ and\
  \bibinfo {author} {\bibfnamefont {K.~S.}\ \bibnamefont {Korolev}},\
  }\bibfield  {title} {\bibinfo {title} {Slow expanders invade by forming
  dented fronts in microbial colonies},\ }\href
  {https://doi.org/10.1073/pnas.2108653119} {\bibfield  {journal} {\bibinfo
  {journal} {Proceedings of the National Academy of Sciences}\ }\textbf
  {\bibinfo {volume} {119}},\ \bibinfo {pages} {e2108653119} (\bibinfo {year}
  {2022})}\BibitemShut {NoStop}%
\bibitem [{\citenamefont {Pigolotti}\ \emph {et~al.}(2007)\citenamefont
  {Pigolotti}, \citenamefont {López},\ and\ \citenamefont
  {Hernández-García}}]{pigolotti_species_2007}%
  \BibitemOpen
  \bibfield  {author} {\bibinfo {author} {\bibfnamefont {S.}~\bibnamefont
  {Pigolotti}}, \bibinfo {author} {\bibfnamefont {C.}~\bibnamefont {López}},\
  and\ \bibinfo {author} {\bibfnamefont {E.}~\bibnamefont
  {Hernández-García}},\ }\bibfield  {title} {\bibinfo {title} {Species
  {Clustering} in {Competitive} {Lotka}-{Volterra} {Models}},\ }\href
  {https://doi.org/10.1103/PhysRevLett.98.258101} {\bibfield  {journal}
  {\bibinfo  {journal} {Physical Review Letters}\ }\textbf {\bibinfo {volume}
  {98}},\ \bibinfo {pages} {258101} (\bibinfo {year} {2007})}\BibitemShut
  {NoStop}%
\bibitem [{\citenamefont {Andreguetto~Maciel}\ and\ \citenamefont
  {Martinez-Garcia}(2021)}]{andreguetto_maciel_enhanced_2021}%
  \BibitemOpen
  \bibfield  {author} {\bibinfo {author} {\bibfnamefont {G.}~\bibnamefont
  {Andreguetto~Maciel}}\ and\ \bibinfo {author} {\bibfnamefont
  {R.}~\bibnamefont {Martinez-Garcia}},\ }\bibfield  {title} {\bibinfo {title}
  {Enhanced species coexistence in {Lotka}-{Volterra} competition models due to
  nonlocal interactions},\ }\href {https://doi.org/10.1016/j.jtbi.2021.110872}
  {\bibfield  {journal} {\bibinfo  {journal} {Journal of Theoretical Biology}\
  }\textbf {\bibinfo {volume} {530}},\ \bibinfo {pages} {110872} (\bibinfo
  {year} {2021})}\BibitemShut {NoStop}%
\bibitem [{\citenamefont {Zelnik}\ \emph
  {et~al.}(2024{\natexlab{a}})\citenamefont {Zelnik}, \citenamefont {Barbier},
  \citenamefont {Shanafelt}, \citenamefont {Loreau},\ and\ \citenamefont
  {Germain}}]{loreau2024Linking}%
  \BibitemOpen
  \bibfield  {author} {\bibinfo {author} {\bibfnamefont {Y.~R.}\ \bibnamefont
  {Zelnik}}, \bibinfo {author} {\bibfnamefont {M.}~\bibnamefont {Barbier}},
  \bibinfo {author} {\bibfnamefont {D.~W.}\ \bibnamefont {Shanafelt}}, \bibinfo
  {author} {\bibfnamefont {M.}~\bibnamefont {Loreau}},\ and\ \bibinfo {author}
  {\bibfnamefont {R.~M.}\ \bibnamefont {Germain}},\ }\bibfield  {title}
  {\bibinfo {title} {Linking intrinsic scales of ecological processes to
  characteristic scales of biodiversity and functioning patterns},\ }\href
  {https://doi.org/https://doi.org/10.1111/oik.10514} {\bibfield  {journal}
  {\bibinfo  {journal} {Oikos}\ }\textbf {\bibinfo {volume} {2024}},\ \bibinfo
  {pages} {e10514} (\bibinfo {year} {2024}{\natexlab{a}})}\BibitemShut
  {NoStop}%
\bibitem [{\citenamefont {May}(1972)}]{may_will_1972}%
  \BibitemOpen
  \bibfield  {author} {\bibinfo {author} {\bibfnamefont {R.~M.}\ \bibnamefont
  {May}},\ }\bibfield  {title} {\bibinfo {title} {Will a {Large} {Complex}
  {System} be {Stable}?},\ }\href {https://doi.org/10.1038/238413a0} {\bibfield
   {journal} {\bibinfo  {journal} {Nature}\ }\textbf {\bibinfo {volume}
  {238}},\ \bibinfo {pages} {413} (\bibinfo {year} {1972})}\BibitemShut
  {NoStop}%
\bibitem [{\citenamefont {Bunin}(2017)}]{bunin_ecological_2017}%
  \BibitemOpen
  \bibfield  {author} {\bibinfo {author} {\bibfnamefont {G.}~\bibnamefont
  {Bunin}},\ }\bibfield  {title} {\bibinfo {title} {Ecological communities with
  {Lotka}-{Volterra} dynamics},\ }\href
  {https://doi.org/10.1103/PhysRevE.95.042414} {\bibfield  {journal} {\bibinfo
  {journal} {Physical Review E}\ }\textbf {\bibinfo {volume} {95}},\ \bibinfo
  {pages} {042414} (\bibinfo {year} {2017})}\BibitemShut {NoStop}%
\bibitem [{\citenamefont {Galla}(2018)}]{galla_dynamically_2018}%
  \BibitemOpen
  \bibfield  {author} {\bibinfo {author} {\bibfnamefont {T.}~\bibnamefont
  {Galla}},\ }\bibfield  {title} {\bibinfo {title} {Dynamically evolved
  community size and stability of random {Lotka}-{Volterra} ecosystems},\
  }\href {https://doi.org/10.1209/0295-5075/123/48004} {\bibfield  {journal}
  {\bibinfo  {journal} {Europhysics Letters}\ }\textbf {\bibinfo {volume}
  {123}},\ \bibinfo {pages} {48004} (\bibinfo {year} {2018})}\BibitemShut
  {NoStop}%
\bibitem [{\citenamefont {Altieri}\ \emph
  {et~al.}(2021{\natexlab{a}})\citenamefont {Altieri}, \citenamefont {Roy},
  \citenamefont {Cammarota},\ and\ \citenamefont
  {Biroli}}]{altieri_glassy2021}%
  \BibitemOpen
  \bibfield  {author} {\bibinfo {author} {\bibfnamefont {A.}~\bibnamefont
  {Altieri}}, \bibinfo {author} {\bibfnamefont {F.}~\bibnamefont {Roy}},
  \bibinfo {author} {\bibfnamefont {C.}~\bibnamefont {Cammarota}},\ and\
  \bibinfo {author} {\bibfnamefont {G.}~\bibnamefont {Biroli}},\ }\bibfield
  {title} {\bibinfo {title} {Properties of equilibria and glassy phases of the
  random lotka-volterra model with demographic noise},\ }\href
  {https://doi.org/10.1103/PhysRevLett.126.258301} {\bibfield  {journal}
  {\bibinfo  {journal} {Phys. Rev. Lett.}\ }\textbf {\bibinfo {volume} {126}},\
  \bibinfo {pages} {258301} (\bibinfo {year} {2021}{\natexlab{a}})}\BibitemShut
  {NoStop}%
\bibitem [{\citenamefont {Aguirre-L{\'o}pez}(2024)}]{aguirre2024heterogeneous}%
  \BibitemOpen
  \bibfield  {author} {\bibinfo {author} {\bibfnamefont {F.}~\bibnamefont
  {Aguirre-L{\'o}pez}},\ }\bibfield  {title} {\bibinfo {title} {Heterogeneous
  mean-field analysis of the generalized lotka-volterra model on a network},\
  }\href@noop {} {\bibfield  {journal} {\bibinfo  {journal} {arXiv preprint
  arXiv:2404.11164}\ } (\bibinfo {year} {2024})}\BibitemShut {NoStop}%
\bibitem [{\citenamefont {Poley}\ \emph {et~al.}(2024)\citenamefont {Poley},
  \citenamefont {Galla},\ and\ \citenamefont {Baron}}]{poley2024interaction}%
  \BibitemOpen
  \bibfield  {author} {\bibinfo {author} {\bibfnamefont {L.}~\bibnamefont
  {Poley}}, \bibinfo {author} {\bibfnamefont {T.}~\bibnamefont {Galla}},\ and\
  \bibinfo {author} {\bibfnamefont {J.~W.}\ \bibnamefont {Baron}},\ }\bibfield
  {title} {\bibinfo {title} {Interaction networks in persistent lotka-volterra
  communities},\ }\href@noop {} {\bibfield  {journal} {\bibinfo  {journal}
  {arXiv preprint arXiv:2404.08600}\ } (\bibinfo {year} {2024})}\BibitemShut
  {NoStop}%
\bibitem [{\citenamefont {Park}\ \emph {et~al.}(2024)\citenamefont {Park},
  \citenamefont {Lee}, \citenamefont {Lee},\ and\ \citenamefont
  {Park}}]{park2024incorporating}%
  \BibitemOpen
  \bibfield  {author} {\bibinfo {author} {\bibfnamefont {J.~I.}\ \bibnamefont
  {Park}}, \bibinfo {author} {\bibfnamefont {D.-S.}\ \bibnamefont {Lee}},
  \bibinfo {author} {\bibfnamefont {S.~H.}\ \bibnamefont {Lee}},\ and\ \bibinfo
  {author} {\bibfnamefont {H.~J.}\ \bibnamefont {Park}},\ }\bibfield  {title}
  {\bibinfo {title} {Incorporating heterogeneous interactions for ecological
  biodiversity},\ }\href@noop {} {\bibfield  {journal} {\bibinfo  {journal}
  {Physical Review Letters}\ }\textbf {\bibinfo {volume} {133}},\ \bibinfo
  {pages} {198402} (\bibinfo {year} {2024})}\BibitemShut {NoStop}%
\bibitem [{\citenamefont {Baron}\ \emph {et~al.}(2023)\citenamefont {Baron},
  \citenamefont {Jewell}, \citenamefont {Ryder},\ and\ \citenamefont
  {Galla}}]{baron_breakdown_2023}%
  \BibitemOpen
  \bibfield  {author} {\bibinfo {author} {\bibfnamefont {J.~W.}\ \bibnamefont
  {Baron}}, \bibinfo {author} {\bibfnamefont {T.~J.}\ \bibnamefont {Jewell}},
  \bibinfo {author} {\bibfnamefont {C.}~\bibnamefont {Ryder}},\ and\ \bibinfo
  {author} {\bibfnamefont {T.}~\bibnamefont {Galla}},\ }\bibfield  {title}
  {\bibinfo {title} {Breakdown of {Random}-{Matrix} {Universality} in
  {Persistent} {Lotka}-{Volterra} {Communities}},\ }\href
  {https://doi.org/10.1103/PhysRevLett.130.137401} {\bibfield  {journal}
  {\bibinfo  {journal} {Physical Review Letters}\ }\textbf {\bibinfo {volume}
  {130}},\ \bibinfo {pages} {137401} (\bibinfo {year} {2023})}\BibitemShut
  {NoStop}%
\bibitem [{\citenamefont {Arnoulx~de Pirey}\ and\ \citenamefont
  {Bunin}(2023)}]{arnoulx_de_pirey_aging_prl2023}%
  \BibitemOpen
  \bibfield  {author} {\bibinfo {author} {\bibfnamefont {T.}~\bibnamefont
  {Arnoulx~de Pirey}}\ and\ \bibinfo {author} {\bibfnamefont {G.}~\bibnamefont
  {Bunin}},\ }\bibfield  {title} {\bibinfo {title} {Aging by
  {Near}-{Extinctions} in {Many}-{Variable} {Interacting} {Populations}},\
  }\href {https://doi.org/10.1103/PhysRevLett.130.098401} {\bibfield  {journal}
  {\bibinfo  {journal} {Physical Review Letters}\ }\textbf {\bibinfo {volume}
  {130}},\ \bibinfo {pages} {098401} (\bibinfo {year} {2023})}\BibitemShut
  {NoStop}%
\bibitem [{\citenamefont {Suweis}\ \emph {et~al.}(2024)\citenamefont {Suweis},
  \citenamefont {Ferraro}, \citenamefont {Grilletta}, \citenamefont {Azaele},\
  and\ \citenamefont {Maritan}}]{suweis2024generalized}%
  \BibitemOpen
  \bibfield  {author} {\bibinfo {author} {\bibfnamefont {S.}~\bibnamefont
  {Suweis}}, \bibinfo {author} {\bibfnamefont {F.}~\bibnamefont {Ferraro}},
  \bibinfo {author} {\bibfnamefont {C.}~\bibnamefont {Grilletta}}, \bibinfo
  {author} {\bibfnamefont {S.}~\bibnamefont {Azaele}},\ and\ \bibinfo {author}
  {\bibfnamefont {A.}~\bibnamefont {Maritan}},\ }\bibfield  {title} {\bibinfo
  {title} {Generalized lotka-volterra systems with time correlated stochastic
  interactions},\ }\href@noop {} {\bibfield  {journal} {\bibinfo  {journal}
  {Physical Review Letters}\ }\textbf {\bibinfo {volume} {133}},\ \bibinfo
  {pages} {167101} (\bibinfo {year} {2024})}\BibitemShut {NoStop}%
\bibitem [{\citenamefont {Arnoulx~de Pirey}\ and\ \citenamefont
  {Bunin}(2024)}]{arnoulx_de_pirey_prx_2024}%
  \BibitemOpen
  \bibfield  {author} {\bibinfo {author} {\bibfnamefont {T.}~\bibnamefont
  {Arnoulx~de Pirey}}\ and\ \bibinfo {author} {\bibfnamefont {G.}~\bibnamefont
  {Bunin}},\ }\bibfield  {title} {\bibinfo {title} {Many-{Species} {Ecological}
  {Fluctuations} as a {Jump} {Process} from the {Brink} of {Extinction}},\
  }\href {https://doi.org/10.1103/PhysRevX.14.011037} {\bibfield  {journal}
  {\bibinfo  {journal} {Physical Review X}\ }\textbf {\bibinfo {volume} {14}},\
  \bibinfo {pages} {011037} (\bibinfo {year} {2024})}\BibitemShut {NoStop}%
\bibitem [{\citenamefont {Garcia~Lorenzana}\ \emph {et~al.}(2024)\citenamefont
  {Garcia~Lorenzana}, \citenamefont {Altieri},\ and\ \citenamefont
  {Biroli}}]{lorenzana2024}%
  \BibitemOpen
  \bibfield  {author} {\bibinfo {author} {\bibfnamefont {G.}~\bibnamefont
  {Garcia~Lorenzana}}, \bibinfo {author} {\bibfnamefont {A.}~\bibnamefont
  {Altieri}},\ and\ \bibinfo {author} {\bibfnamefont {G.}~\bibnamefont
  {Biroli}},\ }\bibfield  {title} {\bibinfo {title} {Interactions and migration
  rescuing ecological diversity},\ }\href
  {https://doi.org/10.1103/PRXLife.2.013014} {\bibfield  {journal} {\bibinfo
  {journal} {PRX Life}\ }\textbf {\bibinfo {volume} {2}},\ \bibinfo {pages}
  {013014} (\bibinfo {year} {2024})}\BibitemShut {NoStop}%
\bibitem [{\citenamefont {Denk}\ and\ \citenamefont
  {Hallatschek}(2024)}]{denk_tipping_2024}%
  \BibitemOpen
  \bibfield  {author} {\bibinfo {author} {\bibfnamefont {J.}~\bibnamefont
  {Denk}}\ and\ \bibinfo {author} {\bibfnamefont {O.}~\bibnamefont
  {Hallatschek}},\ }\bibfield  {title} {\bibinfo {title} {Tipping points emerge
  from weak mutualism in metacommunities},\ }\href
  {https://doi.org/10.1371/journal.pcbi.1011899} {\bibfield  {journal}
  {\bibinfo  {journal} {PLOS Computational Biology}\ }\textbf {\bibinfo
  {volume} {20}},\ \bibinfo {pages} {1} (\bibinfo {year} {2024})},\ \bibinfo
  {note} {publisher: Public Library of Science}\BibitemShut {NoStop}%
\bibitem [{\citenamefont {Olmeda}\ and\ \citenamefont
  {Rulands}(2023)}]{olmeda2023long}%
  \BibitemOpen
  \bibfield  {author} {\bibinfo {author} {\bibfnamefont {F.}~\bibnamefont
  {Olmeda}}\ and\ \bibinfo {author} {\bibfnamefont {S.}~\bibnamefont
  {Rulands}},\ }\bibfield  {title} {\bibinfo {title} {Long-range interactions
  and disorder facilitate pattern formation in spatial complex systems},\
  }\href@noop {} {\bibfield  {journal} {\bibinfo  {journal} {arXiv preprint
  arXiv:2303.12611}\ } (\bibinfo {year} {2023})}\BibitemShut {NoStop}%
\bibitem [{\citenamefont {Padmanabha}\ \emph {et~al.}(2024)\citenamefont
  {Padmanabha}, \citenamefont {Nicoletti}, \citenamefont {Bernardi},
  \citenamefont {Suweis}, \citenamefont {Azaele}, \citenamefont {Rinaldo},\
  and\ \citenamefont {Maritan}}]{maritan_pnas2024}%
  \BibitemOpen
  \bibfield  {author} {\bibinfo {author} {\bibfnamefont {P.}~\bibnamefont
  {Padmanabha}}, \bibinfo {author} {\bibfnamefont {G.}~\bibnamefont
  {Nicoletti}}, \bibinfo {author} {\bibfnamefont {D.}~\bibnamefont {Bernardi}},
  \bibinfo {author} {\bibfnamefont {S.}~\bibnamefont {Suweis}}, \bibinfo
  {author} {\bibfnamefont {S.}~\bibnamefont {Azaele}}, \bibinfo {author}
  {\bibfnamefont {A.}~\bibnamefont {Rinaldo}},\ and\ \bibinfo {author}
  {\bibfnamefont {A.}~\bibnamefont {Maritan}},\ }\bibfield  {title} {\bibinfo
  {title} {Landscape and environmental heterogeneity support coexistence in
  competitive metacommunities},\ }\href
  {https://doi.org/10.1073/pnas.2410932121} {\bibfield  {journal} {\bibinfo
  {journal} {Proceedings of the National Academy of Sciences}\ }\textbf
  {\bibinfo {volume} {121}},\ \bibinfo {pages} {e2410932121} (\bibinfo {year}
  {2024})}\BibitemShut {NoStop}%
\bibitem [{\citenamefont {Galla}(2006)}]{galla_random_2006}%
  \BibitemOpen
  \bibfield  {author} {\bibinfo {author} {\bibfnamefont {T.}~\bibnamefont
  {Galla}},\ }\bibfield  {title} {\bibinfo {title} {Random replicators with
  asymmetric couplings},\ }\href {https://doi.org/10.1088/0305-4470/39/15/001}
  {\bibfield  {journal} {\bibinfo  {journal} {Journal of Physics A:
  Mathematical and General}\ }\textbf {\bibinfo {volume} {39}},\ \bibinfo
  {pages} {3853} (\bibinfo {year} {2006})}\BibitemShut {NoStop}%
\bibitem [{\citenamefont {Altieri}\ \emph
  {et~al.}(2021{\natexlab{b}})\citenamefont {Altieri}, \citenamefont {Roy},
  \citenamefont {Cammarota},\ and\ \citenamefont
  {Biroli}}]{altieri_properties_2021}%
  \BibitemOpen
  \bibfield  {author} {\bibinfo {author} {\bibfnamefont {A.}~\bibnamefont
  {Altieri}}, \bibinfo {author} {\bibfnamefont {F.}~\bibnamefont {Roy}},
  \bibinfo {author} {\bibfnamefont {C.}~\bibnamefont {Cammarota}},\ and\
  \bibinfo {author} {\bibfnamefont {G.}~\bibnamefont {Biroli}},\ }\bibfield
  {title} {\bibinfo {title} {Properties of {Equilibria} and {Glassy} {Phases}
  of the {Random} {Lotka}-{Volterra} {Model} with {Demographic} {Noise}},\
  }\href {https://doi.org/10.1103/PhysRevLett.126.258301} {\bibfield  {journal}
  {\bibinfo  {journal} {Physical Review Letters}\ }\textbf {\bibinfo {volume}
  {126}},\ \bibinfo {pages} {258301} (\bibinfo {year}
  {2021}{\natexlab{b}})}\BibitemShut {NoStop}%
\bibitem [{\citenamefont {Garnier-Brun}\ \emph {et~al.}(2021)\citenamefont
  {Garnier-Brun}, \citenamefont {Benzaquen}, \citenamefont {Ciliberti},\ and\
  \citenamefont {Bouchaud}}]{garnier2021new}%
  \BibitemOpen
  \bibfield  {author} {\bibinfo {author} {\bibfnamefont {J.}~\bibnamefont
  {Garnier-Brun}}, \bibinfo {author} {\bibfnamefont {M.}~\bibnamefont
  {Benzaquen}}, \bibinfo {author} {\bibfnamefont {S.}~\bibnamefont
  {Ciliberti}},\ and\ \bibinfo {author} {\bibfnamefont {J.-P.}\ \bibnamefont
  {Bouchaud}},\ }\bibfield  {title} {\bibinfo {title} {A new spin on optimal
  portfolios and ecological equilibria},\ }\href@noop {} {\bibfield  {journal}
  {\bibinfo  {journal} {Journal of Statistical Mechanics: Theory and
  Experiment}\ }\textbf {\bibinfo {volume} {2021}},\ \bibinfo {pages} {093408}
  (\bibinfo {year} {2021})}\BibitemShut {NoStop}%
\bibitem [{\citenamefont {Ahmadian}\ \emph {et~al.}(2015)\citenamefont
  {Ahmadian}, \citenamefont {Fumarola},\ and\ \citenamefont
  {Miller}}]{ahmadian2015}%
  \BibitemOpen
  \bibfield  {author} {\bibinfo {author} {\bibfnamefont {Y.}~\bibnamefont
  {Ahmadian}}, \bibinfo {author} {\bibfnamefont {F.}~\bibnamefont {Fumarola}},\
  and\ \bibinfo {author} {\bibfnamefont {K.~D.}\ \bibnamefont {Miller}},\
  }\bibfield  {title} {\bibinfo {title} {Properties of networks with partially
  structured and partially random connectivity},\ }\href
  {https://doi.org/10.1103/PhysRevE.91.012820} {\bibfield  {journal} {\bibinfo
  {journal} {Phys. Rev. E}\ }\textbf {\bibinfo {volume} {91}},\ \bibinfo
  {pages} {012820} (\bibinfo {year} {2015})}\BibitemShut {NoStop}%
\bibitem [{\citenamefont {Stone}(2018)}]{stone_feasibility_2018}%
  \BibitemOpen
  \bibfield  {author} {\bibinfo {author} {\bibfnamefont {L.}~\bibnamefont
  {Stone}},\ }\bibfield  {title} {\bibinfo {title} {The feasibility and
  stability of large complex biological networks: a random matrix approach},\
  }\href {https://doi.org/10.1038/s41598-018-26486-2} {\bibfield  {journal}
  {\bibinfo  {journal} {Scientific Reports}\ }\textbf {\bibinfo {volume} {8}},\
  \bibinfo {pages} {8246} (\bibinfo {year} {2018})}\BibitemShut {NoStop}%
\bibitem [{\citenamefont {Tuck}(2006)}]{tuck_positivity_2006}%
  \BibitemOpen
  \bibfield  {author} {\bibinfo {author} {\bibfnamefont {E.}~\bibnamefont
  {Tuck}},\ }\bibfield  {title} {\bibinfo {title} {On {Positivity} of {Fourier}
  {Transforms}},\ }\href {https://doi.org/10.1017/S0004972700047511} {\bibfield
   {journal} {\bibinfo  {journal} {Bulletin of the Australian Mathematical
  Society}\ }\textbf {\bibinfo {volume} {74}},\ \bibinfo {pages} {133}
  (\bibinfo {year} {2006})}\BibitemShut {NoStop}%
\bibitem [{\citenamefont {Giraud}\ and\ \citenamefont
  {Peschanski}(2014)}]{giraud_positivity_2014}%
  \BibitemOpen
  \bibfield  {author} {\bibinfo {author} {\bibfnamefont {B.~G.}\ \bibnamefont
  {Giraud}}\ and\ \bibinfo {author} {\bibfnamefont {R.}~\bibnamefont
  {Peschanski}},\ }\href {http://arxiv.org/abs/1405.3155} {\emph {\bibinfo
  {title} {On the positivity of {Fourier} transforms}}},\ \bibinfo {type}
  {Tech. Rep.}\ \bibinfo {number} {arXiv:1405.3155}\ (\bibinfo  {institution}
  {arXiv},\ \bibinfo {year} {2014})\ \bibinfo {note} {arXiv:1405.3155 [hep-ph,
  physics:hep-th, physics:math-ph, physics:nucl-th] type: article}\BibitemShut
  {NoStop}%
\bibitem [{\citenamefont {Turing}(1990)}]{turing1990chemical}%
  \BibitemOpen
  \bibfield  {author} {\bibinfo {author} {\bibfnamefont {A.~M.}\ \bibnamefont
  {Turing}},\ }\bibfield  {title} {\bibinfo {title} {The chemical basis of
  morphogenesis},\ }\href@noop {} {\bibfield  {journal} {\bibinfo  {journal}
  {Bulletin of mathematical biology}\ }\textbf {\bibinfo {volume} {52}},\
  \bibinfo {pages} {153} (\bibinfo {year} {1990})}\BibitemShut {NoStop}%
\bibitem [{\citenamefont {Berestycki}\ \emph {et~al.}(2009)\citenamefont
  {Berestycki}, \citenamefont {Nadin}, \citenamefont {Perthame},\ and\
  \citenamefont {Ryzhik}}]{berestycki_non-local_2009}%
  \BibitemOpen
  \bibfield  {author} {\bibinfo {author} {\bibfnamefont {H.}~\bibnamefont
  {Berestycki}}, \bibinfo {author} {\bibfnamefont {G.}~\bibnamefont {Nadin}},
  \bibinfo {author} {\bibfnamefont {B.}~\bibnamefont {Perthame}},\ and\
  \bibinfo {author} {\bibfnamefont {L.}~\bibnamefont {Ryzhik}},\ }\bibfield
  {title} {\bibinfo {title} {The non-local {Fisher}–{KPP} equation:
  travelling waves and steady states},\ }\href
  {https://doi.org/10.1088/0951-7715/22/12/002} {\bibfield  {journal} {\bibinfo
   {journal} {Nonlinearity}\ }\textbf {\bibinfo {volume} {22}},\ \bibinfo
  {pages} {2813} (\bibinfo {year} {2009})}\BibitemShut {NoStop}%
\bibitem [{\citenamefont {Achleitner}\ and\ \citenamefont
  {Kuehn}(2015)}]{achleitner_bounded_2015}%
  \BibitemOpen
  \bibfield  {author} {\bibinfo {author} {\bibfnamefont {F.}~\bibnamefont
  {Achleitner}}\ and\ \bibinfo {author} {\bibfnamefont {C.}~\bibnamefont
  {Kuehn}},\ }\bibfield  {title} {\bibinfo {title} {On bounded positive
  stationary solutions for a nonlocal {Fisher}–{KPP} equation},\ }\href
  {https://doi.org/10.1016/j.na.2014.09.004} {\bibfield  {journal} {\bibinfo
  {journal} {Nonlinear Analysis: Theory, Methods \& Applications}\ }\textbf
  {\bibinfo {volume} {112}},\ \bibinfo {pages} {15} (\bibinfo {year}
  {2015})}\BibitemShut {NoStop}%
\bibitem [{\citenamefont {Kuehn}\ and\ \citenamefont
  {Throm}(2018)}]{kuehn_validity_2017}%
  \BibitemOpen
  \bibfield  {author} {\bibinfo {author} {\bibfnamefont {C.}~\bibnamefont
  {Kuehn}}\ and\ \bibinfo {author} {\bibfnamefont {S.}~\bibnamefont {Throm}},\
  }\bibfield  {title} {\bibinfo {title} {Validity of amplitude equations for
  nonlocal nonlinearities},\ }\href {https://doi.org/10.1063/1.4993112}
  {\bibfield  {journal} {\bibinfo  {journal} {Journal of Mathematical Physics}\
  }\textbf {\bibinfo {volume} {59}},\ \bibinfo {pages} {071510} (\bibinfo
  {year} {2018})}\BibitemShut {NoStop}%
\bibitem [{\citenamefont {Faye}\ and\ \citenamefont
  {Holzer}(2015)}]{faye_modulated_2015}%
  \BibitemOpen
  \bibfield  {author} {\bibinfo {author} {\bibfnamefont {G.}~\bibnamefont
  {Faye}}\ and\ \bibinfo {author} {\bibfnamefont {M.}~\bibnamefont {Holzer}},\
  }\bibfield  {title} {\bibinfo {title} {Modulated traveling fronts for a
  nonlocal {Fisher}-{KPP} equation: {A} dynamical systems approach},\ }\href
  {https://doi.org/10.1016/j.jde.2014.12.006} {\bibfield  {journal} {\bibinfo
  {journal} {Journal of Differential Equations}\ }\textbf {\bibinfo {volume}
  {258}},\ \bibinfo {pages} {2257} (\bibinfo {year} {2015})}\BibitemShut
  {NoStop}%
\bibitem [{\citenamefont {Baik}\ \emph {et~al.}(2005)\citenamefont {Baik},
  \citenamefont {Arous},\ and\ \citenamefont
  {P{\'e}ch{\'e}}}]{baik_ben-arous_2005}%
  \BibitemOpen
  \bibfield  {author} {\bibinfo {author} {\bibfnamefont {J.}~\bibnamefont
  {Baik}}, \bibinfo {author} {\bibfnamefont {G.~B.}\ \bibnamefont {Arous}},\
  and\ \bibinfo {author} {\bibfnamefont {S.}~\bibnamefont {P{\'e}ch{\'e}}},\
  }\bibfield  {title} {\bibinfo {title} {{Phase transition of the largest
  eigenvalue for nonnull complex sample covariance matrices}},\ }\href
  {https://doi.org/10.1214/009117905000000233} {\bibfield  {journal} {\bibinfo
  {journal} {The Annals of Probability}\ }\textbf {\bibinfo {volume} {33}},\
  \bibinfo {pages} {1643 } (\bibinfo {year} {2005})}\BibitemShut {NoStop}%
\bibitem [{\citenamefont {Monmeyran}\ \emph {et~al.}(2021)\citenamefont
  {Monmeyran}, \citenamefont {Benyoussef}, \citenamefont {Thomen},
  \citenamefont {Dahmane}, \citenamefont {Baliarda}, \citenamefont {Jules},
  \citenamefont {Aymerich},\ and\ \citenamefont {Henry}}]{monmeyran2021four}%
  \BibitemOpen
  \bibfield  {author} {\bibinfo {author} {\bibfnamefont {A.}~\bibnamefont
  {Monmeyran}}, \bibinfo {author} {\bibfnamefont {W.}~\bibnamefont
  {Benyoussef}}, \bibinfo {author} {\bibfnamefont {P.}~\bibnamefont {Thomen}},
  \bibinfo {author} {\bibfnamefont {N.}~\bibnamefont {Dahmane}}, \bibinfo
  {author} {\bibfnamefont {A.}~\bibnamefont {Baliarda}}, \bibinfo {author}
  {\bibfnamefont {M.}~\bibnamefont {Jules}}, \bibinfo {author} {\bibfnamefont
  {S.}~\bibnamefont {Aymerich}},\ and\ \bibinfo {author} {\bibfnamefont
  {N.}~\bibnamefont {Henry}},\ }\bibfield  {title} {\bibinfo {title} {Four
  species of bacteria deterministically assemble to form a stable biofilm in a
  millifluidic channel},\ }\href@noop {} {\bibfield  {journal} {\bibinfo
  {journal} {npj Biofilms and Microbiomes}\ }\textbf {\bibinfo {volume} {7}},\
  \bibinfo {pages} {64} (\bibinfo {year} {2021})}\BibitemShut {NoStop}%
\bibitem [{\citenamefont {Hu}\ \emph {et~al.}(2022)\citenamefont {Hu},
  \citenamefont {Amor}, \citenamefont {Barbier}, \citenamefont {Bunin},\ and\
  \citenamefont {Gore}}]{hu_gore2022}%
  \BibitemOpen
  \bibfield  {author} {\bibinfo {author} {\bibfnamefont {J.}~\bibnamefont
  {Hu}}, \bibinfo {author} {\bibfnamefont {D.~R.}\ \bibnamefont {Amor}},
  \bibinfo {author} {\bibfnamefont {M.}~\bibnamefont {Barbier}}, \bibinfo
  {author} {\bibfnamefont {G.}~\bibnamefont {Bunin}},\ and\ \bibinfo {author}
  {\bibfnamefont {J.}~\bibnamefont {Gore}},\ }\bibfield  {title} {\bibinfo
  {title} {Emergent phases of ecological diversity and dynamics mapped in
  microcosms},\ }\href {https://doi.org/10.1126/science.abm7841} {\bibfield
  {journal} {\bibinfo  {journal} {Science}\ }\textbf {\bibinfo {volume}
  {378}},\ \bibinfo {pages} {85} (\bibinfo {year} {2022})}\BibitemShut
  {NoStop}%
\bibitem [{\citenamefont {{Alejandro Martínez-Calvo}}\ \emph
  {et~al.}(2023)\citenamefont {{Alejandro Martínez-Calvo}}, \citenamefont
  {{Carolina Trenado-Yuste}}, \citenamefont {{Hyunseok Lee}}, \citenamefont
  {{Jeff Gore}}, \citenamefont {{Ned S. Wingreen}},\ and\ \citenamefont {{Sujit
  S. Datta}}}]{alejandro_martinez_interfacial_2023}%
  \BibitemOpen
  \bibfield  {author} {\bibinfo {author} {\bibnamefont {{Alejandro
  Martínez-Calvo}}}, \bibinfo {author} {\bibnamefont {{Carolina
  Trenado-Yuste}}}, \bibinfo {author} {\bibnamefont {{Hyunseok Lee}}}, \bibinfo
  {author} {\bibnamefont {{Jeff Gore}}}, \bibinfo {author} {\bibnamefont {{Ned
  S. Wingreen}}},\ and\ \bibinfo {author} {\bibnamefont {{Sujit S. Datta}}},\
  }\bibfield  {title} {\bibinfo {title} {Interfacial morphodynamics of
  proliferating microbial communities},\ }\href
  {https://doi.org/10.1101/2023.10.23.563665} {\bibfield  {journal} {\bibinfo
  {journal} {bioRxiv}\ ,\ \bibinfo {pages} {2023.10.23.563665}} (\bibinfo
  {year} {2023})}\BibitemShut {NoStop}%
\bibitem [{\citenamefont {Curatolo}\ \emph {et~al.}(2020)\citenamefont
  {Curatolo}, \citenamefont {Zhou}, \citenamefont {Zhao}, \citenamefont {Liu},
  \citenamefont {Daerr}, \citenamefont {Tailleur},\ and\ \citenamefont
  {Huang}}]{curatolo2020cooperative}%
  \BibitemOpen
  \bibfield  {author} {\bibinfo {author} {\bibfnamefont {A.}~\bibnamefont
  {Curatolo}}, \bibinfo {author} {\bibfnamefont {N.}~\bibnamefont {Zhou}},
  \bibinfo {author} {\bibfnamefont {Y.}~\bibnamefont {Zhao}}, \bibinfo {author}
  {\bibfnamefont {C.}~\bibnamefont {Liu}}, \bibinfo {author} {\bibfnamefont
  {A.}~\bibnamefont {Daerr}}, \bibinfo {author} {\bibfnamefont
  {J.}~\bibnamefont {Tailleur}},\ and\ \bibinfo {author} {\bibfnamefont
  {J.}~\bibnamefont {Huang}},\ }\bibfield  {title} {\bibinfo {title}
  {Cooperative pattern formation in multi-component bacterial systems through
  reciprocal motility regulation},\ }\href@noop {} {\bibfield  {journal}
  {\bibinfo  {journal} {Nature Physics}\ }\textbf {\bibinfo {volume} {16}},\
  \bibinfo {pages} {1152} (\bibinfo {year} {2020})}\BibitemShut {NoStop}%
\bibitem [{\citenamefont {Bohmann}\ \emph {et~al.}(2014)\citenamefont
  {Bohmann}, \citenamefont {Evans}, \citenamefont {Gilbert}, \citenamefont
  {Carvalho}, \citenamefont {Creer}, \citenamefont {Knapp}, \citenamefont
  {Douglas},\ and\ \citenamefont {De~Bruyn}}]{bohmann2014environmental}%
  \BibitemOpen
  \bibfield  {author} {\bibinfo {author} {\bibfnamefont {K.}~\bibnamefont
  {Bohmann}}, \bibinfo {author} {\bibfnamefont {A.}~\bibnamefont {Evans}},
  \bibinfo {author} {\bibfnamefont {M.~T.~P.}\ \bibnamefont {Gilbert}},
  \bibinfo {author} {\bibfnamefont {G.~R.}\ \bibnamefont {Carvalho}}, \bibinfo
  {author} {\bibfnamefont {S.}~\bibnamefont {Creer}}, \bibinfo {author}
  {\bibfnamefont {M.}~\bibnamefont {Knapp}}, \bibinfo {author} {\bibfnamefont
  {W.~Y.}\ \bibnamefont {Douglas}},\ and\ \bibinfo {author} {\bibfnamefont
  {M.}~\bibnamefont {De~Bruyn}},\ }\bibfield  {title} {\bibinfo {title}
  {Environmental dna for wildlife biology and biodiversity monitoring},\
  }\href@noop {} {\bibfield  {journal} {\bibinfo  {journal} {Trends in ecology
  \& evolution}\ }\textbf {\bibinfo {volume} {29}},\ \bibinfo {pages} {358}
  (\bibinfo {year} {2014})}\BibitemShut {NoStop}%
\bibitem [{\citenamefont {Aglieri}\ \emph {et~al.}(2021)\citenamefont
  {Aglieri}, \citenamefont {Baillie}, \citenamefont {Mariani}, \citenamefont
  {Cattano}, \citenamefont {Cal{\`o}}, \citenamefont {Turco}, \citenamefont
  {Spatafora}, \citenamefont {Di~Franco}, \citenamefont {Di~Lorenzo},
  \citenamefont {Guidetti} \emph {et~al.}}]{aglieri2021environmental}%
  \BibitemOpen
  \bibfield  {author} {\bibinfo {author} {\bibfnamefont {G.}~\bibnamefont
  {Aglieri}}, \bibinfo {author} {\bibfnamefont {C.}~\bibnamefont {Baillie}},
  \bibinfo {author} {\bibfnamefont {S.}~\bibnamefont {Mariani}}, \bibinfo
  {author} {\bibfnamefont {C.}~\bibnamefont {Cattano}}, \bibinfo {author}
  {\bibfnamefont {A.}~\bibnamefont {Cal{\`o}}}, \bibinfo {author}
  {\bibfnamefont {G.}~\bibnamefont {Turco}}, \bibinfo {author} {\bibfnamefont
  {D.}~\bibnamefont {Spatafora}}, \bibinfo {author} {\bibfnamefont
  {A.}~\bibnamefont {Di~Franco}}, \bibinfo {author} {\bibfnamefont
  {M.}~\bibnamefont {Di~Lorenzo}}, \bibinfo {author} {\bibfnamefont
  {P.}~\bibnamefont {Guidetti}}, \emph {et~al.},\ }\bibfield  {title} {\bibinfo
  {title} {Environmental dna effectively captures functional diversity of
  coastal fish communities},\ }\href@noop {} {\bibfield  {journal} {\bibinfo
  {journal} {Molecular Ecology}\ }\textbf {\bibinfo {volume} {30}},\ \bibinfo
  {pages} {3127} (\bibinfo {year} {2021})}\BibitemShut {NoStop}%
\bibitem [{\citenamefont {Zelnik}\ \emph
  {et~al.}(2024{\natexlab{b}})\citenamefont {Zelnik}, \citenamefont {Galiana},
  \citenamefont {Barbier}, \citenamefont {Loreau}, \citenamefont {Galbraith},\
  and\ \citenamefont {Arnoldi}}]{zelnik_how_2024}%
  \BibitemOpen
  \bibfield  {author} {\bibinfo {author} {\bibfnamefont {Y.~R.}\ \bibnamefont
  {Zelnik}}, \bibinfo {author} {\bibfnamefont {N.}~\bibnamefont {Galiana}},
  \bibinfo {author} {\bibfnamefont {M.}~\bibnamefont {Barbier}}, \bibinfo
  {author} {\bibfnamefont {M.}~\bibnamefont {Loreau}}, \bibinfo {author}
  {\bibfnamefont {E.}~\bibnamefont {Galbraith}},\ and\ \bibinfo {author}
  {\bibfnamefont {J.-F.}\ \bibnamefont {Arnoldi}},\ }\bibfield  {title}
  {\bibinfo {title} {How collectively integrated are ecological communities?},\
  }\href {https://doi.org/10.1111/ele.14358} {\bibfield  {journal} {\bibinfo
  {journal} {Ecology Letters}\ }\textbf {\bibinfo {volume} {27}},\ \bibinfo
  {pages} {e14358} (\bibinfo {year} {2024}{\natexlab{b}})}\BibitemShut
  {NoStop}%
\bibitem [{\citenamefont {Opper}\ and\ \citenamefont
  {Diederich}(1992)}]{opper_phase_1992}%
  \BibitemOpen
  \bibfield  {author} {\bibinfo {author} {\bibfnamefont {M.}~\bibnamefont
  {Opper}}\ and\ \bibinfo {author} {\bibfnamefont {S.}~\bibnamefont
  {Diederich}},\ }\bibfield  {title} {\bibinfo {title} {Phase transition and 1/
  f noise in a game dynamical model},\ }\href
  {https://doi.org/10.1103/PhysRevLett.69.1616} {\bibfield  {journal} {\bibinfo
   {journal} {Physical Review Letters}\ }\textbf {\bibinfo {volume} {69}},\
  \bibinfo {pages} {1616} (\bibinfo {year} {1992})}\BibitemShut {NoStop}%
\bibitem [{\citenamefont {Galla}(2024)}]{galla_step_by_step_2024}%
  \BibitemOpen
  \bibfield  {author} {\bibinfo {author} {\bibfnamefont {T.}~\bibnamefont
  {Galla}},\ }\href {https://arxiv.org/abs/2405.14289} {\bibinfo {title}
  {Generating-functional analysis of random lotka-volterra systems: A
  step-by-step guide}} (\bibinfo {year} {2024}),\ \Eprint
  {https://arxiv.org/abs/2405.14289} {arXiv:2405.14289 [cond-mat.dis-nn]}
  \BibitemShut {NoStop}%
\bibitem [{\citenamefont {M{\'e}zard}\ \emph {et~al.}(1987)\citenamefont
  {M{\'e}zard}, \citenamefont {Parisi},\ and\ \citenamefont
  {Virasoro}}]{mezard1987spin}%
  \BibitemOpen
  \bibfield  {author} {\bibinfo {author} {\bibfnamefont {M.}~\bibnamefont
  {M{\'e}zard}}, \bibinfo {author} {\bibfnamefont {G.}~\bibnamefont {Parisi}},\
  and\ \bibinfo {author} {\bibfnamefont {M.~A.}\ \bibnamefont {Virasoro}},\
  }\href@noop {} {\emph {\bibinfo {title} {Spin glass theory and beyond: An
  Introduction to the Replica Method and Its Applications}}},\ Vol.~\bibinfo
  {volume} {9}\ (\bibinfo  {publisher} {World Scientific Publishing Company},\
  \bibinfo {year} {1987})\BibitemShut {NoStop}%
\bibitem [{\citenamefont {Roy}\ \emph {et~al.}(2019)\citenamefont {Roy},
  \citenamefont {Biroli}, \citenamefont {Bunin},\ and\ \citenamefont
  {Cammarota}}]{Roy_2019}%
  \BibitemOpen
  \bibfield  {author} {\bibinfo {author} {\bibfnamefont {F.}~\bibnamefont
  {Roy}}, \bibinfo {author} {\bibfnamefont {G.}~\bibnamefont {Biroli}},
  \bibinfo {author} {\bibfnamefont {G.}~\bibnamefont {Bunin}},\ and\ \bibinfo
  {author} {\bibfnamefont {C.}~\bibnamefont {Cammarota}},\ }\bibfield  {title}
  {\bibinfo {title} {Numerical implementation of dynamical mean field theory
  for disordered systems: application to the lotka–volterra model of
  ecosystems},\ }\href {https://doi.org/10.1088/1751-8121/ab1f32} {\bibfield
  {journal} {\bibinfo  {journal} {Journal of Physics A: Mathematical and
  Theoretical}\ }\textbf {\bibinfo {volume} {52}},\ \bibinfo {pages} {484001}
  (\bibinfo {year} {2019})}\BibitemShut {NoStop}%
\bibitem [{\citenamefont {Baron}(2022)}]{baron2022eigenvalue}%
  \BibitemOpen
  \bibfield  {author} {\bibinfo {author} {\bibfnamefont {J.~W.}\ \bibnamefont
  {Baron}},\ }\bibfield  {title} {\bibinfo {title} {Eigenvalue spectra and
  stability of directed complex networks},\ }\href@noop {} {\bibfield
  {journal} {\bibinfo  {journal} {Physical Review E}\ }\textbf {\bibinfo
  {volume} {106}},\ \bibinfo {pages} {064302} (\bibinfo {year}
  {2022})}\BibitemShut {NoStop}%
\end{thebibliography}%

\end{document}